ED/GEMR/MRT/2018/P1/19

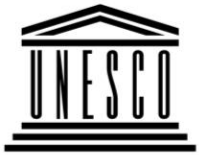
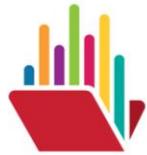

Background paper prepared for the 2019 Global Education Monitoring Report

*Migration, displacement and education:*
*Building bridges, not walls*

# INTERNAL MIGRATION AND EDUCATION:

# A CROSS-NATIONAL COMPARISON

Aude Bernard, Martin Bell and Jim Cooper                    2018

# Contents

# List of Figures

# List of Tables

# Glossary

**Age-Specific Migration Intensities (ASMIs)** are obtained by dividing the number of migrants of age x by the population of age x at the time of census. Since age is recorded at the end of the observation period, when migration is measured over a five-year interval, migrants will have moved on average 2.5 years earlier than the age that is recorded, assuming that migration is evenly distributed over the five-year interval.

**Aggregate Crude Migration Intensity (ACMI)** indicates the number of people who have changed address divided by the population at risk of moving (PAR), expressed as a percentage. Because it measures all changes of address within a country, the ACMI is not affected by the Modifiable Areal Unit Problem (MAUP) and is therefore directly comparable between countries.

**Crude Migration Intensity (CMI)** indicates the number of people who have changed region of residence divided by the population at risk of moving (PAR), expressed as a percentage. The term 'migration intensity', coined by Van Imhoff and Keilman (1991), encompasses both probabilities (when the PAR should be measured as the beginning of the period) and occurrence-exposure rates (for which the PAR should be measured at the middle of the period). In practice, when using census data, the PAR is commonly measured at the end of the observation period.

**Minor and Major Regions** were first distinguished by Rodriguez (2004) to distinguish migration according to two levels of spatial scale. Minor regions correspond to the smallest level of geography available (e.g. counties) and are used to measure short-distance migration, while major regions correspond to a higher level of geography (e.g. states) that encapsulate long-distance migration, such as States or Provinces. The actual number of units in each case still varies widely between countries, so the measures are by no means strictly comparable.

**Modifiable Areal Unit Problem (MAUP),** first clearly described by Openshaw (1984), recognises that the results of any spatial analysis are influenced by the number of units into which an area is divided, and the way the boundaries between them are drawn (Wrigley, Holt, Steel, & Tranmer, 1996). Variations in the spatial scale at which migration is measured therefore affect most measures of migration. A case in point is the Crude Migration Intensity (CMI) which increases with the number of spatial units (Courgeau, 1973) and therefore cannot be used to make direct cross-national comparisons.

**Net Migration Rate (NMR)** is the difference between inflows (total number of people who moved in to a region) and outflows (total number of people who moved out of a region) divided by the population at risk of moving (PAR) multiplied by 100. It measures the degree of net population gain and loss in regions.

**Population at Risk (PAR)** refers to the population at risk of an event (e.g. moving). Choice of the appropriate PAR depends on the type of data being analysed but in practice the population at the end of a census interval is commonly used.

**Spatial scale or spatial framework** refers to the number of spatial units into which countries are divided and often corresponds to administrative regions, such as states and counties. The choice of the spatial framework against which migration is measured is of particular importance as it affects most measures of migration (see MAUP).

**Urban in-migrants** are migrants who moved into urban areas over a given time interval.

# 1. Introduction

Internal migration has now replaced fertility and mortality as the leading agent of demographic change in most countries, and it is the main process shaping patterns of human settlement within and between countries. Migration is widely acknowledged to be integral to the process of human development as it plays a significant role in enhancing educational outcomes. At regional and national levels, internal migration underpins the efficient functioning of the economy by bringing knowledge and skills to the locations where they are needed. At an individual level, migration is essential to economic and social well-being by allowing individuals to pursue their goals and aspirations, including pursuing further study.

It is the multi-dimensional nature of migration that underlines its significance in the process of human development. Unlike other demographic events, such as birth and death, migration is a repetitive process involving varying distance, duration, origins and destinations. Human mobility extends in the spatial domain from local travel to international migration, and in the temporal dimension from short-term stays to permanent relocations. Classification and measurement of such phenomena is inevitably complex (Bell et al., 2002), which has severely hindered progress in comparative research, with very few large-scale cross-national comparisons of migration. In recent years, however, there have been significant methodological advances (Courgeau, Muhidin, & Bell, 2013; Rees et al., 2016) and the emergence of global repositories of internal migration data (Bell, Bernard, Ueffing, & Charles-Edwards, 2014; Minnesota Population Center, 2011), which has given rise to a series of innovative papers under the umbrella of the Internal Migration Around the GlobE (IMAGE) project[1]. This work has made significant headway toward quantifying and understanding the extent of cross-national variations in levels of migration (Bell et al., 2015a), age patterns (Bernard, Bell, & Charles-Edwards, 2014b), distance moved (Stillwell et al., 2016) and population redistribution (Rees et al., 2016), but education and its relationship with migration have not been explicitly considered.

The linkages between migration and education have been explored in a separate line of inquiry that has predominantly focused on country-specific analyses as to the ways in which migration affects educational outcomes and how educational attainment affects migration behaviour. A recurrent theme has been the educational selectivity of migrants, which in turn leads to an increase of human capital in some regions, primarily cities, at the expense of others. Questions have long been raised as to the links between education and migration in response to educational expansion (Gould, 1982; Long, 1973) but have not yet been fully answered because of the absence, until recently, of adequate data for comparative analysis of migration.

In this paper, we bring these two separate strands of research together to systematically explore links between internal migration and education across a global sample of 57 countries at various stages of development, using data drawn from the IPUMS database maintained and made publicly available by the University of Minnesota.[2] The organising framework for our analysis is based around two general classifications. The first distinguishes countries based on their levels of human development, divided into four main classes – low, medium, high and very high. The second classification is spatial and differentiates five broad continental regions: Africa, Asia, Europe, Latin America and the Caribbean, and North America. Some components of the analysis focus on a subset of countries, which have been selected to encompass countries from across the development spectrum and world regions.

---

[1] https://imageproject.com.au/
[2] https://international.ipums.org/international/

In terms of substantive analysis, we seek to compare countries with respect to six broad sets of questions:

1. How does the level of migration vary between countries and by level of educational attainment? Are migrants predominantly selected from highly educated groups? Does the educational selectivity of migrants hold across countries from different regions and at different levels of development? How do these variations affect the educational composition of migration flows?
2. How selective is migration with respect to age? While it is well established that young adults are the most mobile group, there is increasing evidence of variations across countries in the ages at which people move. Do these variations reflect differences in the levels of development? How do migration age patterns vary by levels of educational attainment?
3. How do reasons for moving among adults vary across countries? Does the level and age pattern of education-related migration vary across countries and by sex?
4. To what extent are different educational groups over- or under-represented in particular sets of flows between rural and urban areas? How does the educational attainment of migrants moving to urban areas compare with that of urban and rural stayers? Is the degree of selectivity of urban migrants conditioned by the overall level of schooling in a country?
5. How does migration to urban areas play out in terms of migrant education over time? Do migrants who have resided in urban areas for longer periods of time display higher levels of educational attainment?
6. To what extent does migration redistribute populations across the settlement system? Do the patterns of population redistribution vary with level of development and level of educational attainment?

The paper starts with a review of previous studies on migration and education in Section 2 before discussing obstacles to comparative analysis of migration and presenting the data and methods used in Section 3. Sections 4 to 9 set out the results from our work by addressing each set of questions in turn. Section 10 summaries the findings, discusses how the links between migration and education vary with levels of development and provides recommendations for future research.

# 2. Research on the Links between Migration and Education

Migration has long been recognised as playing an important role in enhancing educational outcomes at both an individual and national level. To date, however, most literature has focused on international migration (OECD, 2017). The links between internal migration and education have received far less attention, even though migration within countries is numerically much more significant than international migration. The latest global estimates put the number of internal migrants living outside their region of birth at 763 million people, which is more than three times the number of international migrants (Bell & Charles-Edwards, 2013). The linkages between migration and education are complex and multifaceted and there are several channels through which migration, both within and between countries, can affect educational outcomes. First, education facilitates migration by lowering the costs and barriers to moving and increasing economic returns - wages in particular - to migration, and a result migrants tend to exhibit a high level of education compared to the general population. Secondly, moving away from home can provide an opportunity for migrants to acquire new skills either through work or study, and in some countries a sizeable proportion of young adults move to pursue further education. At the same time, migration can have a negative effect on the school enrolment of migrants of school age and of children left behind. Finally, the redistribution of human capital through migration can change the stock and composition of skills in both regions of origin and destination. The subsequent paragraphs review each of these topics in turn.

It is now well-established that migrants generally have higher levels of educational attainment than non-migrants, especially over long distances (Long, 1973). A large body of literature dating back to the 1970s has examined the educational selectivity of migrants, with results pointing broadly to a positive effect of educational attainment on the likelihood of migration (Cattaneo, 2007; Greenwood, 1975; Plane & Rogerson, 1994; Williams, 2009) although a few studies have reported a negative relationship (Massey & Espinosa, 1997; Quinn & Rubb, 2005), while others have found no significant association (Adams & Richard, 1993; Curran & Rivero-Fuentes, 2003). Attempts have been made to examine how GPD per capita impacts the migration of groups with different levels of educational attainment, with mixed results (Abel & Muttaraj, 2017; Ginsburg et al., 2016).

In a comparative analysis of migration differentials by education in the United States, Japan, and England, Long (1973: 257) postulated that '*education, however, may be losing some of its ability to predict who migrates in the United States. Recent increases in the number of colleges and the proportion of people going to college have meant that college graduates are becoming a less select group and will probably have less distinctive migration patterns*'. Social theories have subsequently hypothesized that the strength of the relationship between education and migration weakens over time as education expands (Gould, 1982), which suggests that the degree of educational selectivity of migration is likely to vary across countries as a result of differences in levels of educational attainment. This theoretical position is yet to be thoroughly tested because results from existing studies tend to be specific to particular country settings. While attempts have been made to compare the role of education in the decision to migrate in various countries (Temin, Montgomery, Engebretsen, & Barker, 2013), evidence is fragmented. Robust quantification of cross-national variations in the educational selectivity of migrants has been hindered by the lack of comparable data. Recent methodological developments (Courgeau et al., 2013; Rees et al., 2016), combined with the assembly of global repositories of internal migration data (Bell et al., 2014; Minnesota Population Center, 2011), now facilitate such an endeavour.

One recent strand of mobility research has explored changes in the selectivity of migrants, and more broadly in migration behaviour, by focusing on events and transitions in the individual life-course (Kulu & Milewski, 2007). Drawing on the emergence of longitudinal survey data and advances in longitudinal data analysis, the life-course approach emphasises the role of key transitions to adulthood in triggering moves within countries, leading in turn to a concentration of research on migration activity at young adult ages. Four key transitions are generally recognised relating to education, the labour market, union formation and child-bearing. For young adults, expanding schooling sometimes means moving to another region. Education-related migration is well-documented across more developed nations, where it accounts for a significant proportion of moves at young adult ages (Baryla & Dotterweich, 2001; Delaney, Bernard, Harmon, & Ryan, 2007; Faggian, McCann, & Sheppard, 2007; Holdsworth, 2009; Mills, 2006). There are, however, important variations from one country to another in the level and age structure of education-related migration, which relate to differences in the structure of educational markets and in the spatial distribution of opportunities within a country, the presence or not of a cultural impetus to move away to pursue further education (Bernard, Bell, & Charles-Edwards, 2016) and the degree of institutionalisation of the transition between school and work (Heinz, 1999). Education-related migration has rarely been considered explicitly in less developed countries where economic motives are seen as the dominant reason for moving among young adults (Temin et al., 2013).

A distinctive stream of literature, mainly based on qualitative research, has endeavoured to identify the barriers which inhibit young adults from moving for educational purposes in less developed countries, pointing particularly to cost and difficulties in accessing public services at the destination in some countries (Temin et al., 2013). This literature also recognises that while moving away from home can provide an opportunity to acquire new skills either through work or formal education, migrants of school age are less likely to be enrolled in school than child stayers (Katz & Redmond, 2010; OECD, 2017; Temin et al., 2013; Williams, 2009). This is the case for migrants moving both between and within countries, although the educational attainment of young migrants depends on family characteristics, and on the origin and destination (Levels & Dronkers, 2008). This problem is particularly pronounced in China, where the *hukou* system puts unregistered child migrants at a severe disadvantage in accessing school (Liang & Chen, 2007; Liang, Guo, & Duan, 2008). Linked to this is the question of schooling opportunities for migrants' children arising from the absence of one or both parents because of migration. Evidence is, however, mixed and results vary depending on country and family structures (Asis, Huang, & Yeoh, 2004; Battistella & Conaco, 1998).

Since migrants generally have higher levels of educational attainment than non-migrants, the stock and composition of skills in both regions of origin and destination varies as a result of migration. The loss of human capital through the emigration of high-skilled individuals, commonly referred to as 'brain drain', is a recurrent theme in both the internal and international migration literature. There is a substantial body of literature concerned with the movement of human capital, which focuses primarily on how individuals with particular skills or levels of educational attainment move between regions, resulting in a concentration of human capital in some regions at the expense of others. As countries urbanise, internal migrants overwhelming move from rural to urban settlements. The redistribution of populations toward urban areas strengthens as the process of urbanisation accelerates, before slowing down at higher levels of urbanisation and human development (Rees et al., 2016). Thus, as urbanisation proceeds, human capital tends to accumulate in urban areas at the expense of rural areas although the extent of this movement varies over time and between countries (Ginsburg et al., 2016). While the pool of skilled workers at the origin may fall in the short-term, return migration and remittances can help raise the stock of human capital in the longer term (Faggian, Rajbhandari, & Dotzel, 2017; OECD, 2017).

In reviewing the literature between migration and education, it is apparent that despite an increasing focus on detailed case studies that provide in-depth analysis of particular aspects of migration processes, there is a need for a global understanding of the links between migration and education to establish the extent to which the educational selectivity of migrants holds across regions and levels of development.

## 3. Data Linking Migration to Education

An essential pre-requisite to conducting cross-national comparisons is a clear understating of the types of internal migration data collected in countries across around the world. Censuses are the most common source of internal migration data, with 142 countries collecting information on internal migration in the United Nations 2000 round of censuses and 106 countries collecting data in the United Nations 2010 round (Bell & Charles-Edwards, 2013). Globally, fifty countries compile internal migration data from a population register or other administrative collection, while 111 countries draw data from some form of survey, foremost among which are the Demographic and Health surveys managed by USAID (70 countries) (Bell et al., 2015b). Differences in the way these three types of sources collect data create a number of impediments for cross-national comparisons of internal migration. Most fundamental of these are differences in the types of data that are captured, the length of the interval over which migration is measured, and the spatial framework employed. Differences are also found in population coverage, temporal comparability and data quality, as well as in the extent of other information collected, such as the characteristics of migrants, duration of residence and the reasons for migration (Bell et al 2002). Considerable care is therefore needed in making comparisons between countries.

Migration is inherently a spatial process, so a fundamental aspect of interest is to determine the types of places migrants are moving to, and from. Most countries record moves that cross regional and local boundaries, and a distinction is commonly made between major and minor administrative units, such as states or provinces, and districts or municipalities. For the purposes of this report, we refer to these respectively as major and minor regions. Although the nomenclature, geography and number of these units differ widely between countries, this classification does permit a broad distinction to be made between long- and short-distance moves. Because geographical classifications tend to be based on administrative boundaries, they are somewhat less helpful in identifying particular types of movement, such as urban-rural migration. While most censuses do distinguish urban or rural residence at the time of the census, this information is rarely collected for previous place of residence.

The nature of previous place of residence was collected in the Demographic and Health surveys (DHS) implemented in many developing countries since the 1970s but, while these data have been used variously in previous analyses (Temin et al., 2013), they suffer a number of limitations for this type of research. First, they are confined to women aged 15 to 49. Secondly, migration is only recorded as a change of locality, with no identification of a particular geographic origin or destination, so moves recorded as occurring to urban areas might well originate in a separate locality in the same conurbation. We assessed the reliability of this data source by comparing estimates of migration over a five year interval based on DHS data with similar information from Censuses for 22 countries. We found a very low correspondence between the two series (r=0.17). Moreover, the well-established positive relationship linking migration rates to levels of human development (Bell et al., 2015a) is reversed according to the DHS data, casting further doubt on their reliability. Because DHS respondents self-assessed the urban status of their previous place of residence, it is unlikely to correspond to the official classification of urban areas used to categorise their current place of residence. In Sri Lanka, for example, 83 per cent of the population currently live in rural areas, but only 10 per cent of DHS respondents reported living in a rural area at age 12. Such data give a very misleading picture of rural-urban migration flows and we conclude that the use of DHS data to analyse the links between migration and education therefore pose a serious risk of misinterpretation.

For these reasons, we elected to confine attention in this report to the analysis of data from Censuses, for those countries where such data are available. The most extensive and accessible source of such data is the Integrated Public Use Microdata Series-International (IPUMS) database maintained by the Minnesota Population Centre at the University of Minnesota. At the time of writing, IPUMS held census micro sample files for 85 countries, of which 67 included migration data, 57 in a format suitable for this report. We focus primarily on data from the 2010 census round (2005-2014). For countries that have not yet released data collected at their most recent census, data from the 2000 census round has been used instead (1994-2005). Countries vary widely in the interval over which migration is measured (Bell et al. 2015) but to maximise the number of countries for which data are available and ensure comparability, we selected countries that measure migration over a five-year interval, captured either directly by comparing place of residence at two points in time (transition data), or derived by combining data on previous place of residence with duration of residence in current location (duration data). Whenever available, we have used migration data measured at two different spatial scales to distinguish between short and long-distance migration. All countries in IPUMS also collected information on level of educational attainment at their censuses, classified as less than primary, primary, secondary and tertiary education. This classification is based on the United Nations standard of six years of primary schooling, three years of lower secondary schooling, and three years of higher secondary schooling. A subset of 29 countries in Africa, Asia and Latin America also collected years of schooling.

The country sample used in this study encompasses all major world regions with migration data for 15 countries in Africa, 14 in Asia, 8 in Europe, 17 in Latin America and 3 in North America. Of these 57 countries, 34 measured migration at two spatial scales. Drawing on a diverse sample of countries enables us to examine how the links between migration and education vary among countries at different levels of human development and in different regional contexts. However, information on several key items of information is inevitably more limited. While all countries collected the urban status of current place of residence, only 13 also distinguished between urban and rural residence prior to migration, which constrains the analysis of rural to urban migration flows. Of these 13 countries, just eight also collected duration of residence in the current locality, which provides a small but valuable window into the way educational attainment changes with increasing length of residence in urban areas. Just ten countries collected reasons for moving.

Appendix table A.1 lists the selected case study countries by region, specifies the name and number of spatial scales at which migration is measured and indicates the availability in the Census of three key pieces of data: duration of residence, urban/rural status of previous place of residence, and reasons for moving. Appendix table A2 reports the level of development, level of urbanisation and mean years of schooling in each country.

We utilise a range of descriptive statistics and migration indicators to analyse the data, including migration intensities, migration flows, correlation and regression. To compensate for the dearth of rural-urban migration data, we also adopt an alternative solution to understanding the way population groups with different levels of education are being redistributed through the urban hierarchy. Drawing on the techniques recently developed by Rees et al. (2016), this employs regional population density as a proxy for the conventional urban/rural classification. The method is explained in Section 7.

## 4. How Does Migration Vary?

### a. Variation between countries

The probability of changing residence changes over time and varies widely between people of differing characteristics, including age, but a growing body of evidence shows that it also differs markedly according to the country in which people live. In seeking to understand the links between migration and education, it is therefore crucial to recognise these underlying spatial variations.

The overall level of internal migration, that is *the propensity to change residence*, is measured by the crude migration intensity (CMI), which is calculated as the number of people who changed place of residence in a given interval[3], divided by the population at risk of moving, and expressed as a percentage. As noted earlier, migration is commonly recorded by reference to the number of people crossing the boundaries between administrative areas, so cross-national comparisons using crude measures are compromised by differences in the number of spatial units from one country to the next. One solution to this issue is to adopt a synthetic measure - the Aggregate Crude Migration Intensity (ACMI) - which encompasses all changes of address within a country. Although few countries collect these data directly, Courgeau et al. (2012) demonstrated how the ACMI can be estimated from data on migration at a range of spatial levels, and Bell et al. (2015) subsequently applied this technique to make estimates for 95 countries encompassing 80 per cent of the global population.

Figure 1 ranks the 61 of these 95 countries which measure migration over a five year interval from highest to lowest migration intensities, revealing marked variations. Migration intensities vary from a low of 5 per cent in India in to a high of 55 per cent in New Zealand, with a global mean of 21 per cent indicating that on average one person in five changed place of residence over a five-year interval. The highest migration levels are found in the four new world countries – Australia, Canada, New Zealand and the United States, together with South Korea, Fiji, Panama, Chile and Switzerland. Intensities below 10 per cent are scattered across Asia (India, North Korea, Nepal, Iraq and the Philippines), Africa (Egypt and Mali) and Latin America (Venezuela), with Spain being the sole European representative at this end of the mobility scale. Clear spatial patterns emerge both within and between regions. In Europe there is a marked spatial gradient of high mobility in the North and the West and lower mobility in the South and East. A similar gradient is apparent in Latin America and the Caribbean trending from low mobility in Mexico and parts of Central America, increasing to moderate levels in Brazil and Argentina, cumulating in high mobility in Chile through the Andean countries of Bolivia Peru and into Paraguay. In much of Asia, the level of internal migration is generally low, especially in South and South East Asia, with low mobility in India and Nepal, grading to moderate levels in Thailand, Vietnam, Indonesia and China to high levels in South Korea and Japan. Africa also shows considerable diversity in migration levels, with high mobility in the West (Cameroon and Senegal), moderate levels in Morocco, South Africa and Guinea and intensities below the global mean elsewhere.

Explanation for these differences has been sought from a number of perspectives, with one common explanation being the level of development. For the 61 countries in Figure 1, Bell et al. (2015) found a strong positive association between migration intensities and economic development measured in terms of GDP per capita (r=0.55) and social development measured in terms of HDI (r=0.50). Scatter plots of the aggregate crude migration intensity against GDP per capita and HDI can be found in Figures A1 and A2 of the Appendix. The association with HDI is particularly instructive, because the HDI itself is a composite measure combining life expectancy, per capita income and education[4]. Subsequent analysis conducted for this report delivered a correlation coefficient of r=0.56 between the ACMI and the Education Index for the 61 countries in Figure 1.

---

[3] For this paper we confine attention to countries that measure migration over a five year interval: see section 3

Variation in migration intensities cannot be explained solely by reference to the level of national development and a number of other explanations have been put forward relating to differences in institutional frameworks (Long 1991), the structure of housing and labour markets (Long, 1991; Van Der Gaag & Van Wissen, 2008) and levels of political and social openness (Bernard, Rowe, Bell, Ueffing, & Charles-Edward, 2017). Differences in national age composition are also important since migration intensities are higher at younger ages (Bell et al., 2015a). Nevertheless, these results clearly suggest that overall levels of education serve to shape the propensity to migrate within a country. By extension, this also underlines the need to take account of the diversity of national contexts when examining the relationship between the two variables.

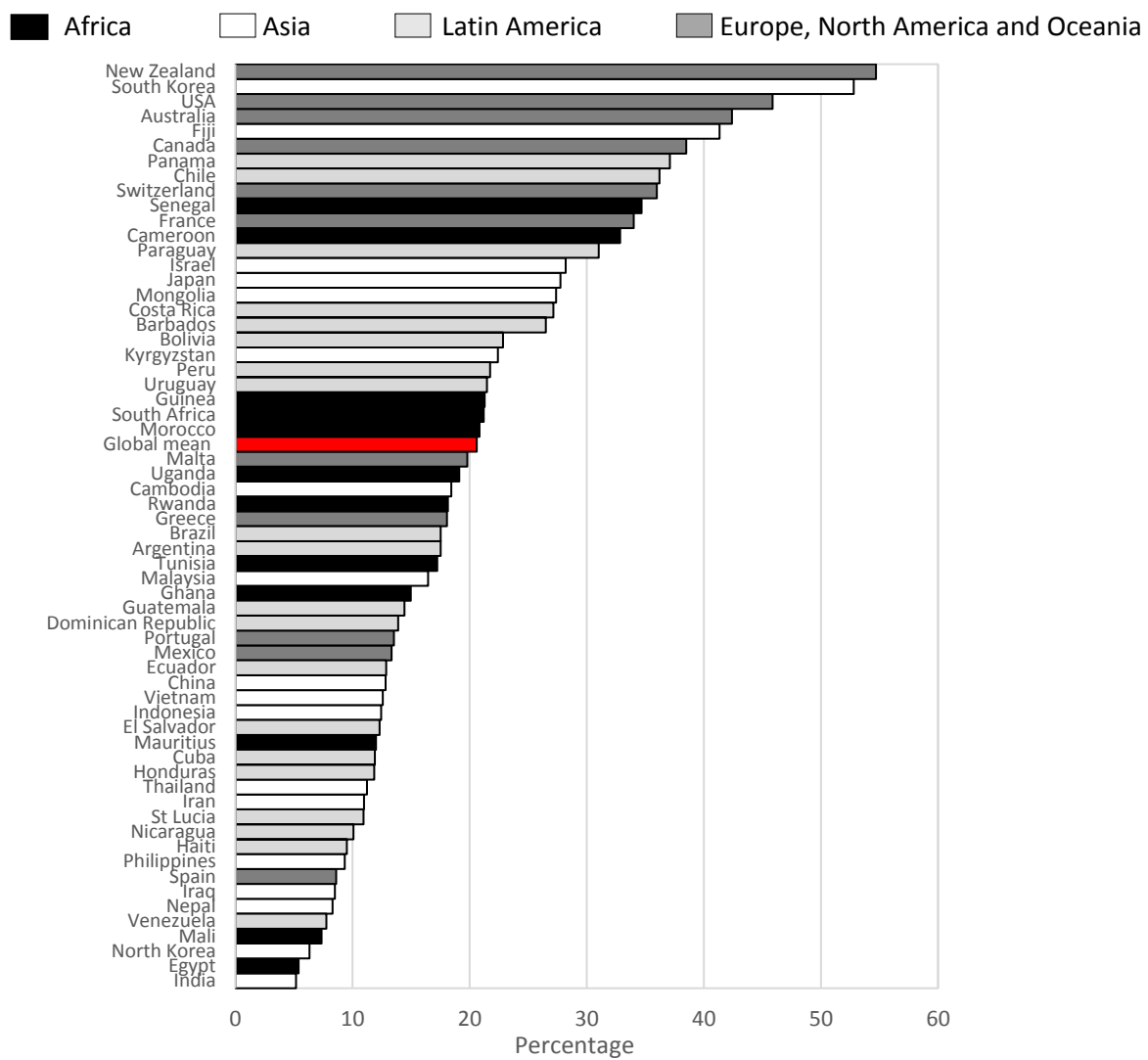

**Figure 1** Five-Year Aggregate Crude Migration Intensities

---

[4] See http://hdr.undp.org/en/data Prior to 2011, which covers most of the data reported here, the education component was measured by the adult literacy rate (with two-thirds weighting) and the combined primary, secondary, and tertiary gross enrolment ratio (with one-third weighting). From 2011, the education index, was computed as a weighted index combining mean years of completed and expected schooling.

## b. Variation by Level of Education

Having established the extent of cross-national variations in overall levels of migrations, we now explore the way migration intensities vary by level of educational attainment. Instead of replicating the approach employed to estimate all changes of address in Figure 1, we utilise the distinction between minor and major regions to produce two sets of Crude Migration Intensities (CMI) that distinguish between short- and long-distance migration. We differentiate levels of intensity by educational attainment by computing CMIs at both spatial scales for four levels of educational attainment (less than primary education, primary education completed, secondary education completed and tertiary education completed) for individuals aged 15 and over. Results for individual countries can be found in Appendix Table A3.

For each country, we computed the ratio of the CMI for individuals with primary education to the CMI for those with less than primary education, and similar ratios for secondary and tertiary education, again using the CMI for individuals with less than primary education as the denominator so that it constitutes the reference population. Thus, a ratio of the CMI for primary education of 2 means that individuals with primary education are twice as likely to move as individuals with no primary education. This approach has the advantage of comparing variations in migration by educational attainment, independent of country differences in migration levels.

Figure 2 reports ratios averaged across each region distinguishing between short and long-distance migration. It reveals that the likelihood of moving increases significantly with the level of education in all world regions. This relationship holds in all countries except for Nicaragua and Guinea for short-distance moves and Iraq for long-distance moves. Globally, compared with individuals with less than primary education, those with primary education are 1.8 times more likely to move between major regions, those with second education 2.8 times and those with tertiary education 3.8 times. The strength of this gradient is slight less pronounced for migration between minor regions, with corresponding ratios of 1.7, 2.5 and 3.2. The positive association between education and migration is particularly pronounced in Asia at both spatial scales and slightly more moderate although still significant in Europe. The gradient is especially strong in Africa for long-distance moves, where individuals with tertiary education are 4.6 times more likely to move than individuals with no primary education, but much weaker in Latin America for both short and long-distance moves, with ratios ranging from 1.3 and 2.1[5].

Several of these continental groupings encompass countries at widely differing positions on the development ladder, so it is important to establish whether this relationship between education and migration holds across the development spectrum. Figure 3 provides unequivocal confirmation. The ratios confirm a consistent increase in mobility with rising levels of education for countries in all four categories of human development[6]. While it would be invidious to make too much of these data, given the modest sample sizes in each category, the ratios suggest that primary and secondary education exert the most pronounced effect in raising levels of migration at the lower end of the development ladder, while tertiary education has its greatest effect among countries in the *very high* HDI category. Because of this differential effect, no simple association is found between the strength of the educational gradient and the level of human development when computed across all countries. Although the consistently positive gradient is striking, the effects of education on the propensity to migrate therefore appear to be more subtle, varied and specific to the individual country context.

---

[5] In comparing these results between spatial scales, recall that short- and long-distance migration data encompass different groups of countries with long-distance migration data available for a larger group of countries in each region. To control for differences in sample composition, we computed mean ratios for the 32 countries that collect migration at both spatial scales, with almost identical results to those reported in Figure 2.

[6] Allocation to the four categories is based on thresholds determined by the United Nations Development Program: see http://hdr.undp.org/en/data.

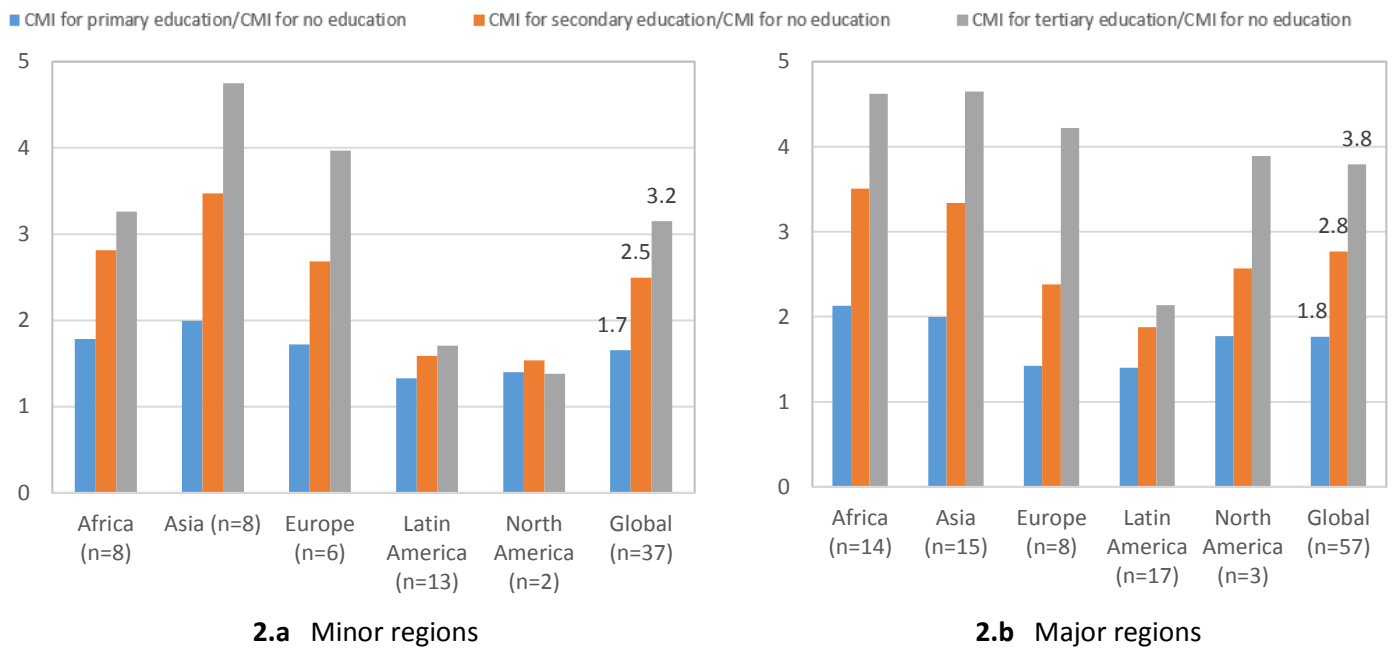

**2.a** Minor regions

**2.b** Major regions

**Figure 2** Five-Year Crude Migration Intensities

*Note: population aged 15 and over*

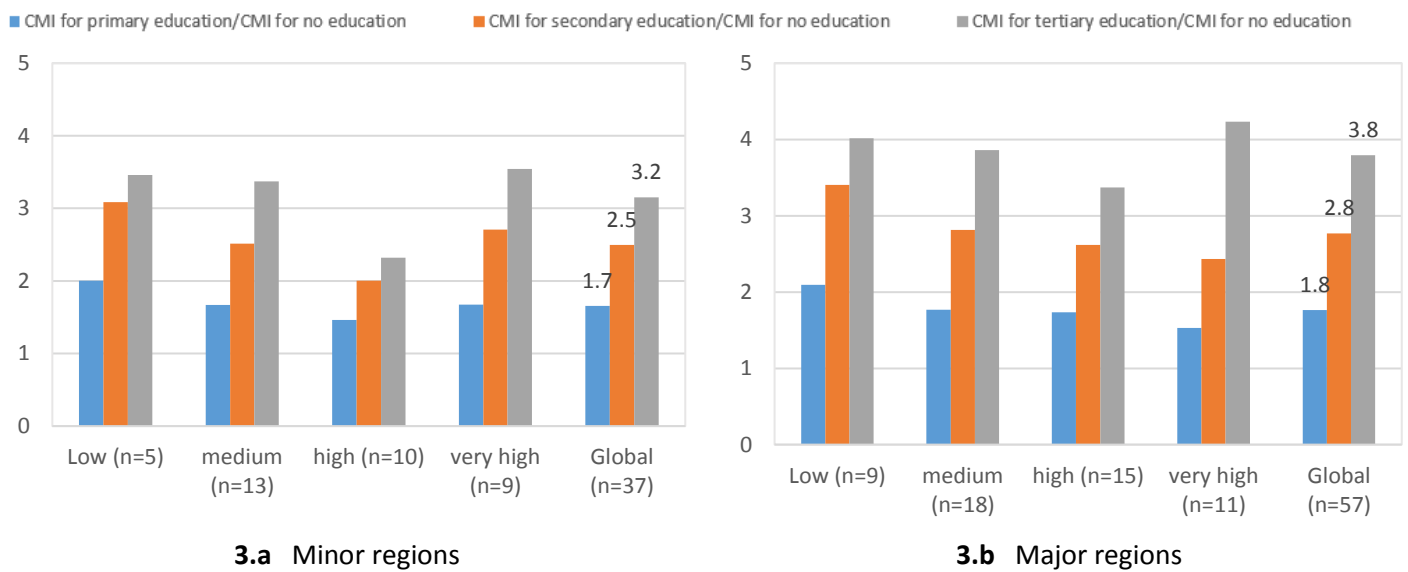

**3.a** Minor regions

**3.b** Major regions

**Figure 3** Mean ratios of CMIs by level of educational attainment and level of human development

Caution is needed in making direct comparison of the size of the bars across regions or levels of HDI in Figures 2 and 3. Because these represent ratios, rather than raw migration intensities, it is not possible to draw conclusions from Figure 3 about the effect of development (as indicated by an increase in HDI) on the actual propensity to migrate among particular educational groups. Reverting to the raw data would not circumvent this problem because the intensities reported are affected by differences in spatial scale: that is in the number of regions between which migration is measured. We are therefore unable to comment directly on the nature of the relationship between migration and development at different levels of education reported elsewhere (Ginsburg et al. 2016; Abel and Muttarak 2017).

### c. The Composition of Migration flows

Although migration rates are systematically larger for those with higher levels of education, the composition of internal migrants is fundamentally dependent on the absolute numbers in each category of educational status. Figure 4 sets out the educational composition of migration streams between major units for a selection of countries in various global regions and at varying stages of development, as measured by the HDI. The interaction between migration rates and populations at risk is abundantly clear. Among countries at the lower end of the development spectrum, high proportions with less than secondary education generate migration profiles that are heavily weighted towards lower levels of educational achievement, despite above average migration intensities among these groups. Conversely, the more advanced educational profiles characteristic of populations in developed countries combine with very high migration intensities to generate migration streams with much larger shares of secondary and tertiary educated people. The net result is to accentuate differences in the educational composition of migration streams between countries with low and high levels of HDI. However, age composition also plays an important role, exerting a moderating effect on these differences, because populations in economically advanced countries are generally older than those in countries earlier in the development process, and so include larger proportions of people at ages where the propensity to migrate is comparatively low.

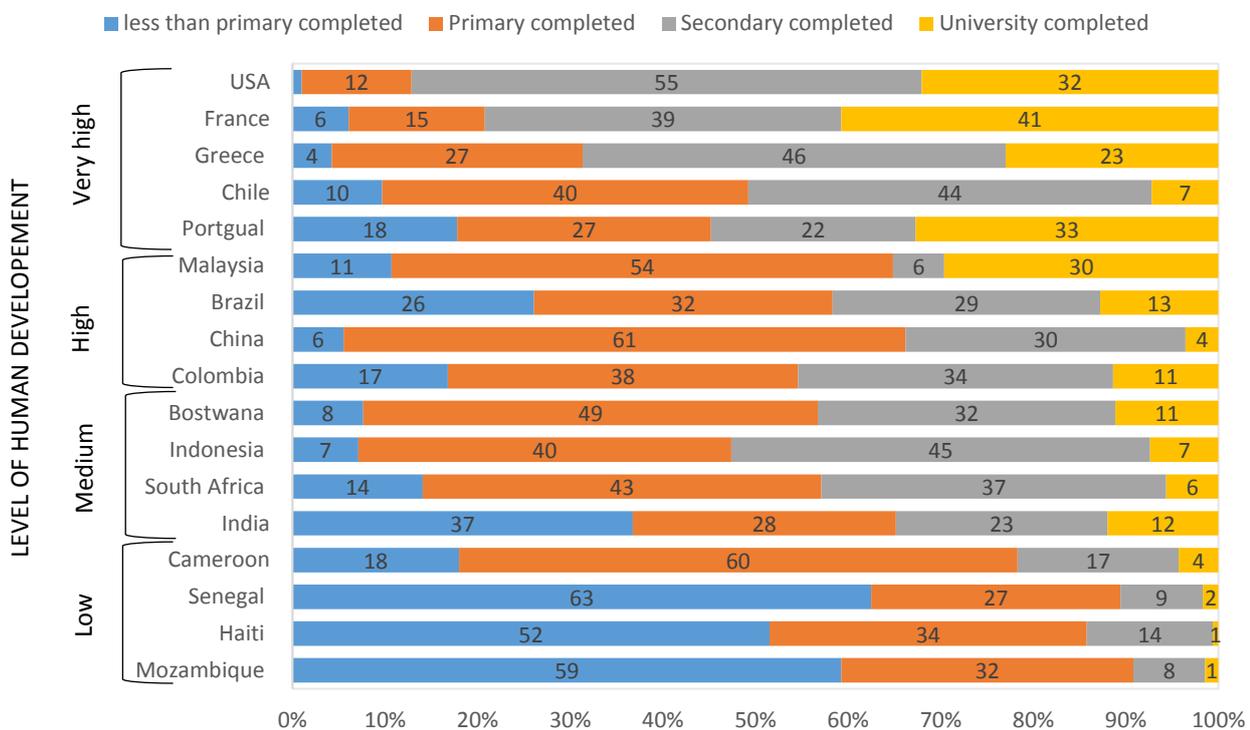

**Figure 4** Educational composition of migrants, selected countries

*Note: population aged 15 and over, migration between major regions, countries ranked in decreasing order of HDI*

In the next section, we examine the relationship between age and migration in detail, but Figure 5 helps clarify the effect of age on the educational composition of migrants between major units for a subset of countries. Countries were selected from across the development spectrum and world regions to provide a concise, yet representative overview. For each selected country, we report levels of educational attainment for five-year age groups encompassing the prime working ages of 15 to 44 years. This allows us to analyse cross-national variations in the educational composition of migrants independently of differences in overall migration levels and migration age patterns. The educational profiles of migrants broadly reflect cross-national differences in levels of educational attainment, but they also reflect country-specific differences in migration rates. The highest proportions of migrants with less than primary education are found in Ghana and Morocco, and this is consistent across all age groups, whereas France and Portugal have the largest proportions with tertiary education. Indonesia, Vietnam, Brazil and Nicaragua display intermediate profiles. Similar findings were found for migration between minor regions (Figure A3 of the Appendix). Analysis by sex shows that this pattern holds for both males and female (Figure A4 and A5), although in African and Asian countries women are on average slightly less educated that men, particularly for older age groups, reflecting the relatively recent increase in educational attainment among these populations.

Superimposed upon these differences, migrants in the 25-29 and 30-34 age groups in all countries tend to display the highest overall levels of educational attainment. In France and Portugal, for example, 61 and 41 per cent of migrants aged 25-29 have completed tertiary education. In Indonesia and Vietnam, the most educated group are a little younger, with the 20-24 age group displaying the largest share with post primary education. For migrants at younger and older ages, the educational attainment of migrants is generally lower than for cohorts in their late twenties and early thirties. Those aged 15-24 may be yet to complete their education, but lower attainment at the older end of the age spectrum almost certainly reflects the relative recency of the rise in educational participation.

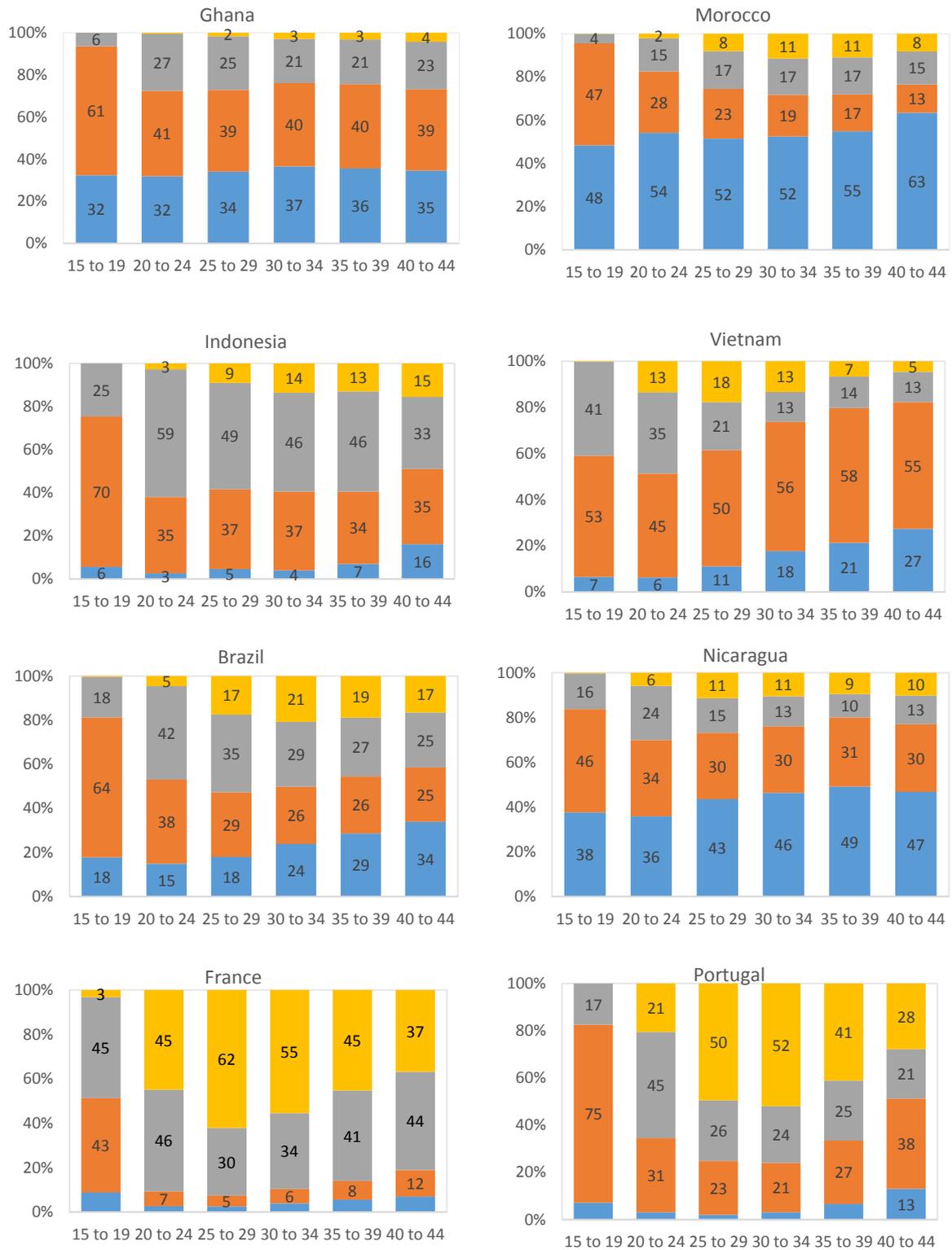

**Figure 5** Percentage distribution of migrants by age and educational attainment

*Note: migration between major regions*

We now turn our attention to the absolute volume of migration flows, which are the product of differences in the base population and in migration intensity. In Africa the largest flow is found in Kenya (3.6 million) followed by Morocco (1.4 million), Cameroon (1.3 million), South Africa and Uganda (both 1.1 million). The smallest flows are seen in Botswana, Senegal and Guinea, with less than 0.3 million migrants. In Kenya, Cameron, South Africa, Uganda, Ghana and Botswana, individuals with primary education constitute the largest flow of migrants while in other African countries it is those with less than primary education that dominate migration flows. Variation in the volume of migration flows is greater in Asia. China reports the largest number of migrants (111 million) followed by India (38.3 million), Indonesia (7.7million) and Vietnam (5.2 million), while flows below 0.5 million are found in Iraq, Kyrgyzstan, Fiji and Armenia. In the majority of Asian countries, the largest flows are of individuals with primary education (China, Vietnam, Iran, Thailand, Iran, Malaysia, Nepal, Fiji), but those with secondary education dominate migration flows in Indonesia, the Philippines, Kyrgyzstan and Armenia. In India, Cambodia and Iraq migration flows are largest among individuals with less than primary education. In Latin America and the Caribbean, the largest migration flows are also found in the most populous countries, namely Brazil (10.1 million), Colombia (2.0 million), Argentina (1.9 million) Chile (1.9 million) and Peru (1.8 million). Conversely, the smallest flows are in countries with small populations such as Nicaragua, Jamaica, El Salvador and Costa Rica. In all countries, individuals with primary education constitute the largest flow of migrants, except for Haiti and Nicaragua where individuals with less than primary education dominate migration flows, and Chile and Peru where individuals with secondary education are the most numerous among migrants, reflecting cross-national differences in levels of educational attainment. In Europe and North America, migration flows range from 19.3 million in the United States to 5.3 million in Mexico and 5.1 million in France, down to 3.6 million in Spain, 2.9 million in Canada and 1.2 million in Switzerland. Less populated countries in Southern and Eastern Europe, including Portugal, Slovenia and Romania report smaller flows below 0.25 million migrants. In all countries, individuals with secondary education constitute the largest flow of migrants, with the exception of France and Portugal where individuals with tertiary education are dominant, and Mexico where individuals with primary education are the most numerous migrants.

## 5. Migration by age

### a. The age profile of migration

Migration is an age-selective process, with young adults being the most mobile group. Irrespective of aggregate levels of mobility the propensity to move typically peaks at young adult ages, then steadily declines with increasing age, rising again among young children who move with their parents, and sometimes around the age of retirement. This broad age profile is replicated, with some variations, at various spatial scales and in a variety of countries (Rogers & Castro, 1981). Migration is most common among young adults because this is the age at which major transitions in the life-course are concentrated. Despite the broad similarities apparent in migration age profiles, there is increasing evidence of systematic variations in the ages at which migration occurs, particular at young adult ages. This is readily apparent in Figure 6, which reports for three countries age-specific migration intensities normalised to unity so that age patterns can be compared independently from differences in the overall level of migration. It shows that migration within China is strongly concentrated in the early 20s, whereas in Brazil and Portugal migration peaks at older ages and is more widely dispersed across the age range.

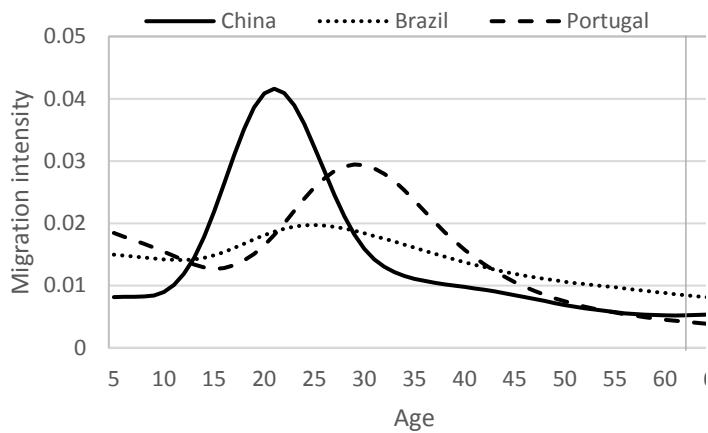

**Figure 6** Age-specific Migration Intensities, selected countries

*Note: Migration data were normalised to sum to unity and smoothed using kernel regression (Bernard & Bell, 2015)*

To elaborate the extent of variations in the age structure of migration in further detail, we constructed the migration age profile for 56 countries. Age-specific migration intensities were obtained by dividing the number of migrants of age x by the population of age x at the time of census. Migration is measured over a five-year interval and since age is recorded at the end of the observation period, migrants will have moved on average 2.5 years earlier than what is recorded, assuming that migration is evenly distributed over the five-year interval. To systematically establish the extent of variation in the age profile of migration for large sample of countries, we use two indicators that have been shown to adequately summarise migration age patterns: the age at which migration peaks, and the intensity of migration at the peak. These two indicators capture two thirds of the inter-country variance in migration age profiles (Bernard, Bell, & Charles-Edwards, 2014a) and have the advantage of having an intrinsic meaning. The age at which migration peaks captures how early in life migration occurs, while the intensity of migration at the peak gauges the degree of concentration of migration activity at young adult ages. Results for all countries can be found in Tables A4.1 and A4.2 of the Appendix.

Using migration between minor regions by way of example[7], Figure 7 plots for 37 countries the age at which migration peaks against the migration intensity at the peak. Each of the small dots represents a country, while the larger dots show regional means, distinguishing Latin America in two groups according to levels of human development. Axes intersect at the global mean and delineate four quadrants. The upper right quadrant gathers European and North American countries where migration peaks in the late twenties and is moderately concentrated around the peak. The lower right quadrant mainly groups Latin American countries at high levels of human development where migration also peaks late but is dispersed across a broader age range. The upper left quadrant features Asian countries where migration peaks in the early twenties and is strongly concentrated around the peak, while the lower left quadrant brings together African countries and Latin American countries at lower levels of development. These results underline the distinctive migration age profiles characteristics of each geographic region. They also highlight the extent of variations among countries within the same region, particularly in Latin America, which appears to be bifurcated into distinctive clusters that reflect differences in levels of human development. Across all countries, the age at which migration peaks returns a correlation coefficient with HDI of 0.59 for migration between minor regions and 0.41[8] for migration between major regions. The difference in correlation coefficients simply reflects the fact that short and long-distance migration encompass different groups of countries. A closer inspection of Figure 7 reveals, however, important variations within regions. Asia is a case in point, with migration peaking at 20.5 years in China and Vietnam compared with 22.5 years in Indonesia and Malaysia and 24.5 in the Philippines.

---

[7] The age profile of migration has been shown be largely scale independent in a way that the age profile of short-distance migration closely matches that of long distance migration (Rogers and Castro 1981; Bell and Muhidin 2009). A total of 33 countries in our sample collect migration both between minor and major regions. The correlation between the age at peak measured at the two spatial scales at is 0.84. It goes up to 0.98 if Chile and Greece are removed, which are the only two countries for which the age profiles of short- and long-distance migration differs

These differences within and between regions, and the association of the age at peak migration with the HDI, can be traced to the fact that the age patterns of migration are fundamentally shaped by the age structure of key events in the life course that mediate the transition to adulthood. Crucial among these transitions are entry into the labour market, union and family formation and, of course, the entry to and completion of education, particularly at post-secondary level (Bernard et al., 2014b). In many Asian societies, the process of becoming an adult is guided by social structures and norms that support early and rapid transitions into adult statuses (Yeung & Alipio, 2013). Thus, the concentration of life-course transitions in early adult life results in turn in a pronounced concentration of migration activity in the early twenties, as shown in Figure 7. In North America and Europe, on the other hand, contemporary practice reflects a much later set of transitions, dispersed more widely across the age spectrum. Key examples include relatively late ages at partnership formation and first birth among women, but this older profile may also reflect more extended involvement in post-secondary education, with the consequential effect of later entry into the labour force.

Primary education is generally undertaken directly from the parental home, and for only a small proportion is secondary education likely to involve a physical relocation, as in the case of children attending boarding school. While residential relation to a desired school catchment occurs in some countries, it mainly involves short-distance moves within the same city. Entry to post-secondary education, on the other hand, has a strong likelihood of requiring a change in residential address and often involves moving over long distances. In some cultures, the start of tertiary education forms no less than a rite of passage, almost invariably involving a move to a share house, or student accommodation often far removed from childhood residence. The UK is a prime example. Elsewhere, as in Australia for example, such moves are much less common, with most tertiary students remaining in the parental home, or at least in the same city (Bell et al. 2002; Bernard et al. 2016). In other countries, especially where opportunities are spatially concentrated, the distribution of educational opportunities may also trigger a relocation to access the desired training. Irrespective of the spatial norms or imperatives at the start of post-secondary study, its completion is commonly associated with a migration triggered by the transition from education into the labour force. Of course, not all moves are triggered by life-course transitions as young adults move in responses to a wide range of opportunities and constraints. Thus, contextual factors may trigger, on occasion, migration directly as in the case of changes in economic conditions (Molloy, Smith, & Wozniak, 2014). The start and end of post-secondary education nevertheless loom large as key forces shaping migration, especially among those in their twenties. We examine the age profile associated with particular reasons for move in more detail in section 6, but it is important first to establish how age interacts with educational achievement in shaping the propensity to move.

---

[8] Excluding Chile and Greece.

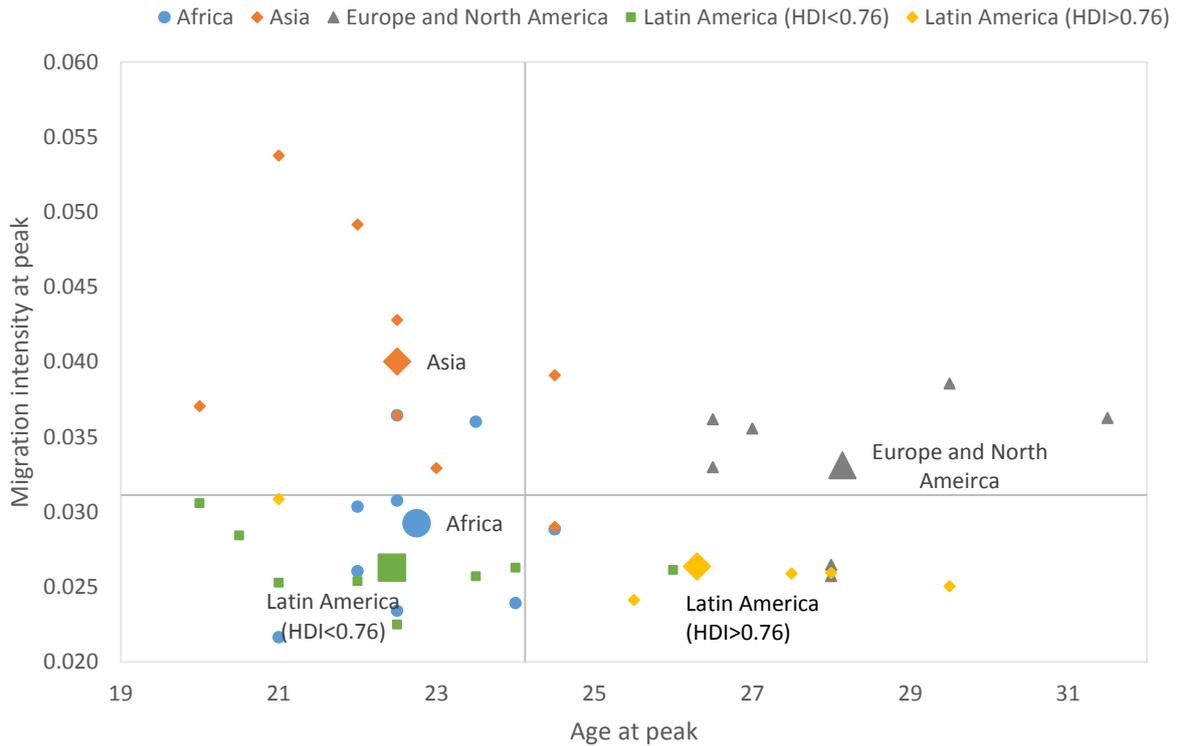

**Figure 7** Age and intensity at peak migration, selected countries and regional means

*Note: axes intersects at the means, regional means* minor regions

### b. Migration by age and educational attainment

As noted above, there is a positive association between migration intensity and educational achievement that holds for all geographic regions and for groups of countries at all levels of HDI. However, the incremental effect of education on mobility varies between countries and levels of educational attainment. The composition of migration flows by level of education also suggested that age-specific effects play an important role in shaping the profile of internal migration. Figure 8 provides more direct evidence. Care is needed in comparing the various countries because the vertical scale differs between graphs, reflecting differences in the sizes of territorial units, but clear regularities are nevertheless apparent.

Senegal, which falls in the group of countries with the lowest HDI, displays a relatively flat age profile in which migration rates at older working ages are only marginally below those recorded for young adults. Despite this, there is evidence of a systematic increase in migration rates with increasing levels of educational attainment and this is sustained across most age groups. Only in the case of tertiary education does the incremental effect disappear, and the numbers in this case are very small, as Figure 4 revealed. Mozambique, which belongs to the same group of countries with low HDI, displays a remarkably similar profile, though with marginally higher rates of movement at the upper end of the educational profile.

Indonesia and Botswana, both countries with Medium levels of HDI, display a distinct and consistent shift compared with their Low HDI counterparts. The difference in intensities between those with only primary, and with less than primary education, has diminished, while intensities for those with tertiary education are more distinctive, rising well above those for people who have completed secondary education. The transition continues in the profiles for Brazil and China, countries with High levels of HDI, in which there is now little difference in migration rates between those with and without primary education. In the case of China, completion of secondary education still delivers a marked stimulus to migration across all age groups, as is apparent in Senegal and Cambodia, but not in Brazil. However, above average mobility rates among the tertiary-educated are evident for both countries and especially pronounced in China. Also notable are the very high rates of movement at ages 15-24 in China among those with secondary or higher education, reflecting the early transition to adulthood noted earlier.

The shift in educational selectivity reaches its zenith in France and the United States at the Very High end of the development spectrum, with profiles in which differences in migration intensity have all but disappeared between those with secondary, primary, or no qualifications, irrespective of age, but strikingly high rates of mobility among those with tertiary qualifications, a feature which is sustained throughout the age profile. Coupled with a highly educated population, the result, as was shown in Figure 4, is an educational profile dominated by people with high levels of educational attainment. Rates of migration among those in the 20-39 age group with post-secondary qualifications are especially pronounced, with up to one third changing residence over a five year period, representing more than double the rates of movement of those with lesser levels of education.

This brief comparison between countries offers qualified support for the proposition advanced by Gould (1982) that the effect of education on mobility diminishes with progress through the educational transition, although-the tertiary educated remain strongly differentiated from the remaining groups. It further suggests that this shift involves a systematic, graded change which shapes both the level and composition of migration. Equally apparent is that it is those at the upper end of the educational profile who play a pivotal role in shaping the overall age profile of migration, particularly in highly developed countries.

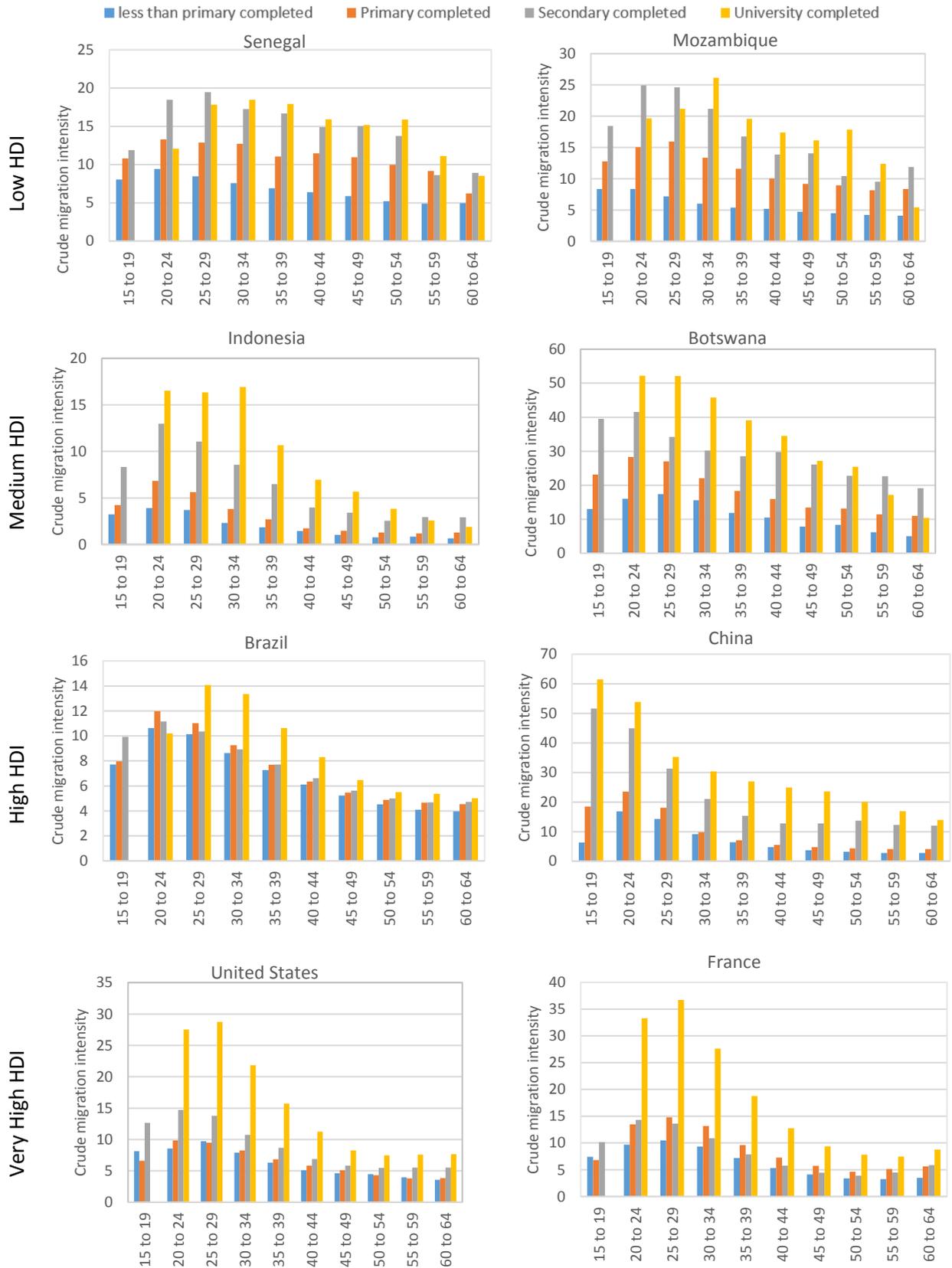

**Figure 8** Crude migration intensities by age and level of educational attainment, selected countries

*Note: population aged 15 and over, migration between major regions. Note the difference between the four graphs in the scale on the vertical axis.*

## 6. Reasons for moving

Migration is a multifaceted phenomenon and people move for a variety of reasons at different times in their lives. Very few countries collect information on reasons for moving at their census and such data are only available for ten countries in the IPUMS[9] collection: Cambodia, China, Colombia, Egypt, India, Indonesia, Iran, Iraq, Mexico and Thailand. Reasons for moving are rarely collected because they face a number of deficiencies, including ex-post rationalisation, event recall bias and the fact that only one main reason is recorded whereas the decision to migrate is a complex, multi-causal process. Despite these limitations, such data can provide useful insights into the motives underpinning migration decisions and how they vary between countries. In this section, we examine the level and age patterns of migration for educational purposes among young adults aged 15 to 24, where most education-related migration occurs, and compare them to other reasons for moving.

Figure 9 reports the distribution of migrants by main reason for moving, distinguishing employment, education, family and marriage, and ranks countries according to the significance of education-related reasons. Because reasons for moving are generally believed to vary with distance, Figure 9 distinguishes between short and long-distance migration, though data at both levels of scale are only available for five of the ten countries. While differences in motivation between the two scales are certainly apparent, the most striking feature of Figure 9 is that education accounts for a minority of moves in all countries, even at the young adult ages which are the focus here. In a majority of countries, employment emerges as the dominant reason for moving between major regions, while family reasons come to the fore in migration over shorter distances. Family reasons also dominate longer distance moves in Iran, Iraq and Egypt, perhaps reflecting cultural norms in predominantly Islamic societies, while in India marriage migration accounts for the largest share of moves, especially over shorter distances.

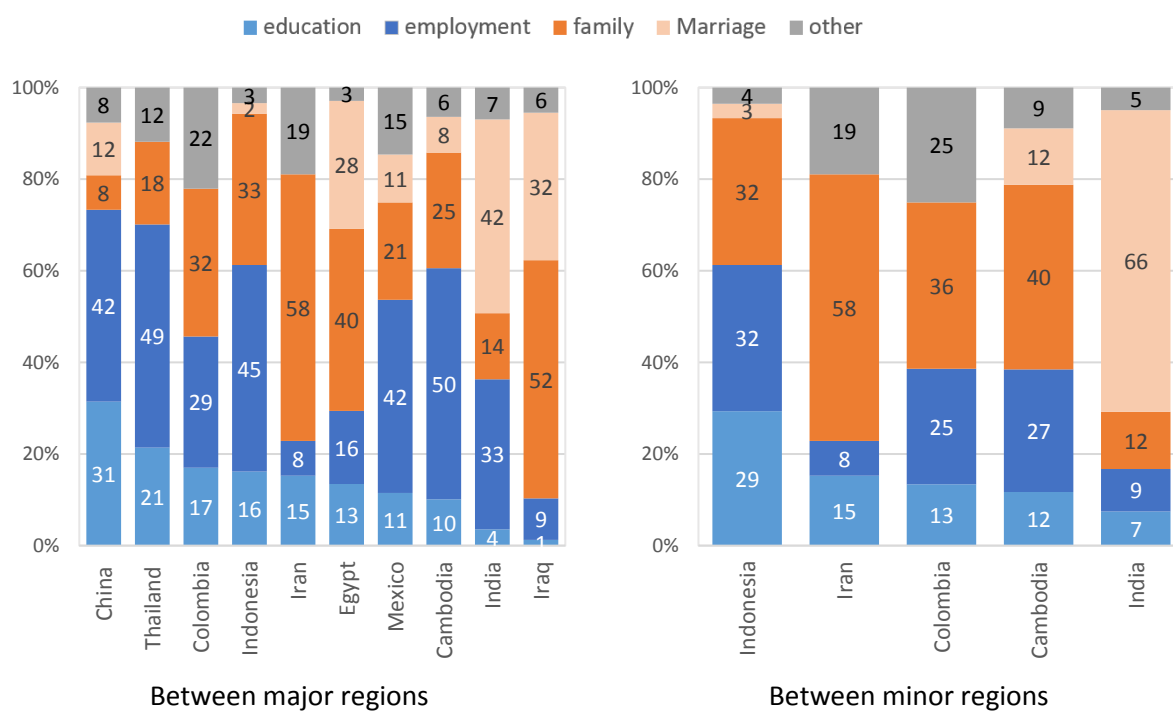

**Figure 9** Reasons for moving, persons aged 15-24, selected countries

---

[9] Three other countries, Armenia, Kyrgyzstan and Uruguay, also asked reasons for moving but did not separately identify education as a motive so they are excluded from the analysis.

Despite its minor role, countries vary widely in regard to the significance of education as a motive for moving. In China, one in three 15-24 year olds migrating over long distances did so for education-related reasons, compared with one in five Thais, one in six Columbians and Indonesians, one in seven Iranians, one in eight Egyptians, one in nine Mexicans and one in ten Cambodians. Education barely featured as a motive for migration in India or Iraq, accounting for less than 5 per cent of all moves at this age. Although the proportions vary, education displays similar levels of significance among short distance migrants.

While education accounts for a minor share of migration overall, there are marked variations between the sexes across countries in its significance as a reason for moving. As shown in Table 1, the proportion of men citing education as their reason for moving consistently exceeds that for women, the sole exception being found in the case of Thailand. Elsewhere, marked differences are found, most prominently in Egypt where education accounted for almost a quarter of long-distance moves among men but just one in twenty in the case of women and in India where education accounted for more than a quarter of short-distance moves among men but just over one in fifty in the case of women. Ratios between the shares for men and women are also high in Iraq, though the proportions moving for education are much smaller. The ratios are somewhat lower in Cambodia, Iran, China and Indonesia, but the absolute differences are more significant, with education accounting for between 5 (Indonesia) and 12 (China) percentage points more of moves among men than among women. Of the ten countries in Table 1, only in Mexico, Colombia and Thailand does education account for similar proportions of movement among men and women. From the results presented here, then, education in many countries is predominantly a male rationale for migration.

| Country | Share of all moves for education (%) | | Ratio of shares Men to Women |
|---|---|---|---|
| | Men | Women | |
| Egypt | 24.5 | 4.5 | 5.5 |
| India | 6.3 | 1.6 | 4.1 |
| Iraq | 2.1 | 0.8 | 2.8 |
| Cambodia | 14.3 | 6.5 | 2.2 |
| Iran | 18.6 | 11.8 | 1.6 |
| China | 38.0 | 26.4 | 1.4 |
| Indonesia | 19.3 | 13.8 | 1.4 |
| Mexico | 12.8 | 10.4 | 1.2 |
| Colombia | 17.2 | 16.9 | 1.0 |
| Thailand | 20.6 | 22.0 | 0.9 |

**Table 1** Sex ratios and percentage share of moves for education between major regions, Persons aged 15-24, selected countries

| Country | Share of all moves for education (%) | | Ratio of shares Men to Women |
|---|---|---|---|
| | Men | Women | |
| India | 26.9 | 2.1 | 12.8 |
| Cambodia | 15.5 | 7.8 | 2.0 |
| Iran | 16.7 | 13.0 | 1.3 |
| Indonesia | 31.4 | 28.0 | 1.1 |
| Colombia | 13.4 | 13.3 | 1.0 |

**Table 2** Sex ratios and percentage share of moves for education between minor regions, Persons aged 15-24, selected countries

Overall variations between countries in the significance of education as a motive for migration can probably be traced to differences in the structure of educational markets and the spatial distribution of educational opportunities within each country, but they are also shaped by cultural and gender-related norms in regard to moving in pursuit of further education (Bernard et al., 2016). The ten countries analysed here are not strongly differentiated in terms of HDI – all lie within the Medium or High categories – nevertheless, it is interesting to note a modest positive correlation between the share of moves attributed to education and the national HDI, with a coefficient of determination (r-squared) of 0.32. This suggests that the role of education as a reason for migration tends to increase in importance as development proceeds.

Figures 10 and 11 provide further insight into education-related moves by tracing the age profile in comparison with other motives, and between countries. Data for selected countries at different levels of development reveal clear and systematic regularities in the shape and timing of reason-specific migration. Taking China by way of example, the country in which educational motives feature most strongly, Figure 10 shows that marriage and education-related migration are more strongly age-graded than employment- and family-related migration, which are dispersed across the life-course. In fact, fully three quarters of education-related moves in China take place between ages 16 and 20, and peak at a significantly earlier age than moves driven by employment or marriage. This result suggests that it is moves for education that contribute strongly to the younger peak in the overall age profile of migration which is characteristic of Asian countries, as reported earlier. The earlier age at the peak among those moving for education compared with other reasons for migration is also clearly apparent in the other countries in Figure 10.

As shown in Figure 11, the profile of education-related migration is remarkably similar across countries, although variations are apparent in both the degree to which such movement is concentrated within a narrow age band, and in the age at which the migration peak is found. While China displays a highly concentrated profile, the profile for Cambodia is more widely dispersed. Subtle variations are also found in the age at the peak, ranging from a low of 18 years in China, to 20 years in Thailand and 22 years in Iran. These differences almost certainly reflect the social, cultural and economic contexts in which education-related moves take place in each country. Unfortunately data on reasons for move are not available to enable comparisons to be made with the age profile of education-related moves for countries at other stages of economic development. However, prior research based on 27 countries representing all major world regions (Bernard et al 2014b) has clearly demonstrated a strong association between the age at which migration peaks and the mean age at completion of formal education.

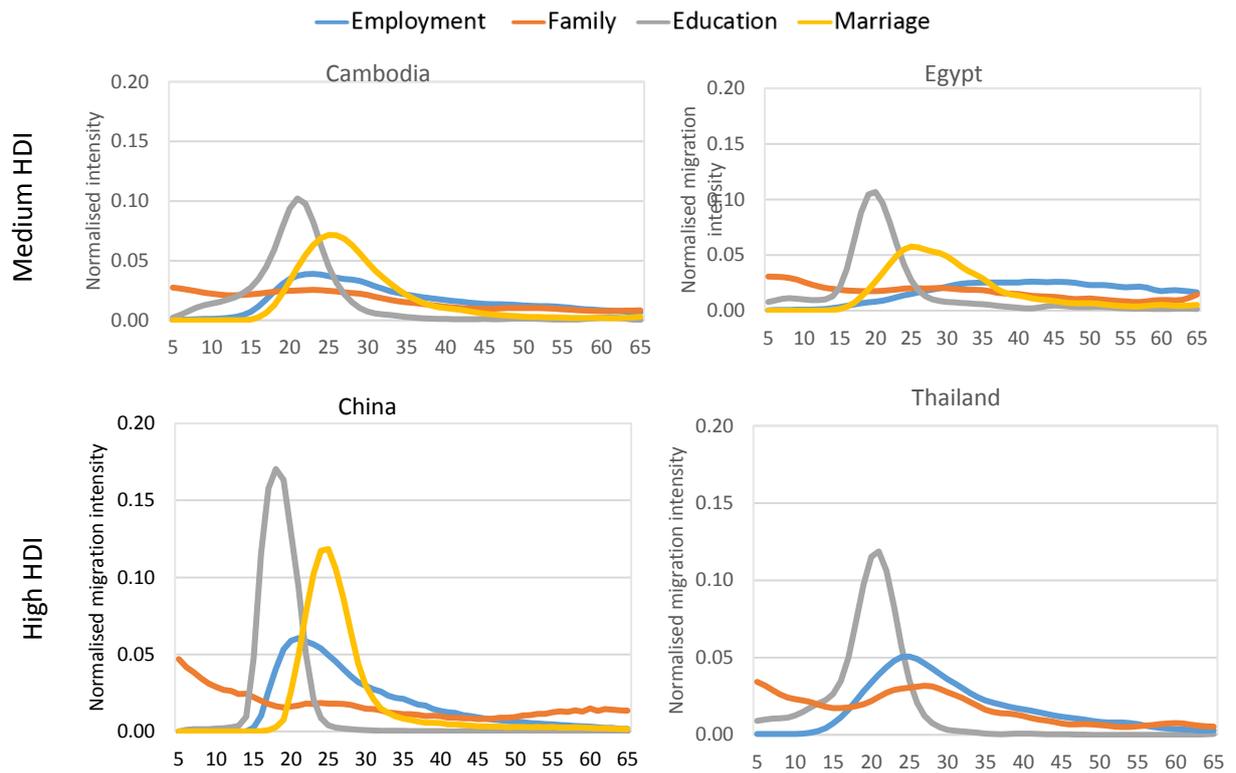

**Figure 10** Age-specific migration intensities by reasons for moving, selected countries, moves between major regions

*Note: Migration data have been smoothed and normalised to sum to unity in order to compare migration age patterns independently from differences in reason-specific migration levels*

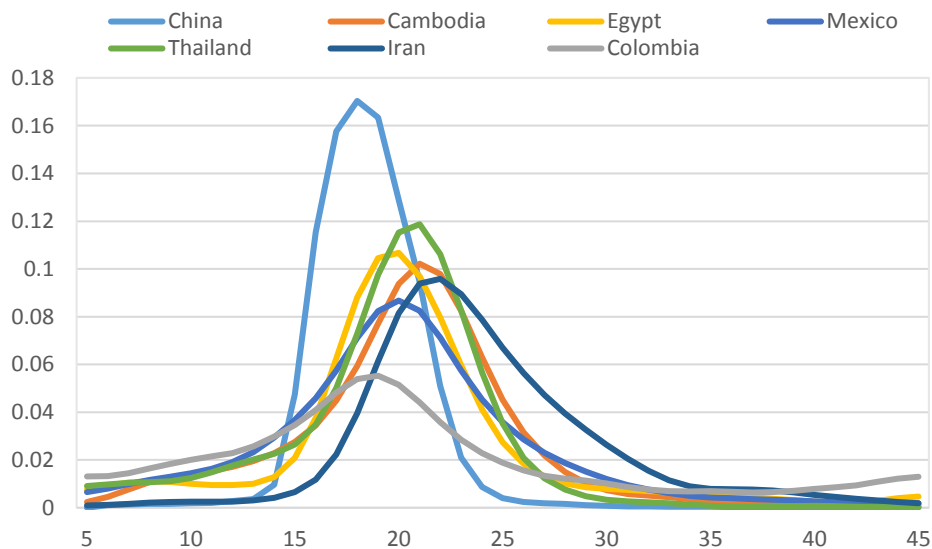

**Figure 11** Age-specific education-related migration intensities, moves between major regions

*Note: Migration data have been smoothed and normalised to sum to unity in order to compare migration age patterns independently from differences in reason-specific migration levels*

# 7. The Direction of Education-related Migration flows

The overall level of mobility tells us about the way in which education facilitates or stimulates migration, but it is the direction of such moves, from village to town or country to city, or in the reverse direction that indicates the way in which migration serves to shape societies. This section reports the direction of flows for segments of the population classified by educational attainment to indicate the composition and relative strength of these movements. Without definitive information tracking individuals through their life course, we cannot assess the precise manner in which migration shapes educational opportunities and outcomes, but for some countries we can document the extent and direction of redistribution of educational groups. As shown in Figure 12, in most countries rural-to-urban migration is numerically less significant than urban to urban moves, or those from one rural area to another, but it is socially significant because it determines the context in which individual lives are embedded, offering new educational and employment opportunities. Together with international migration, it is also the primary process shaping national patterns of human settlement.

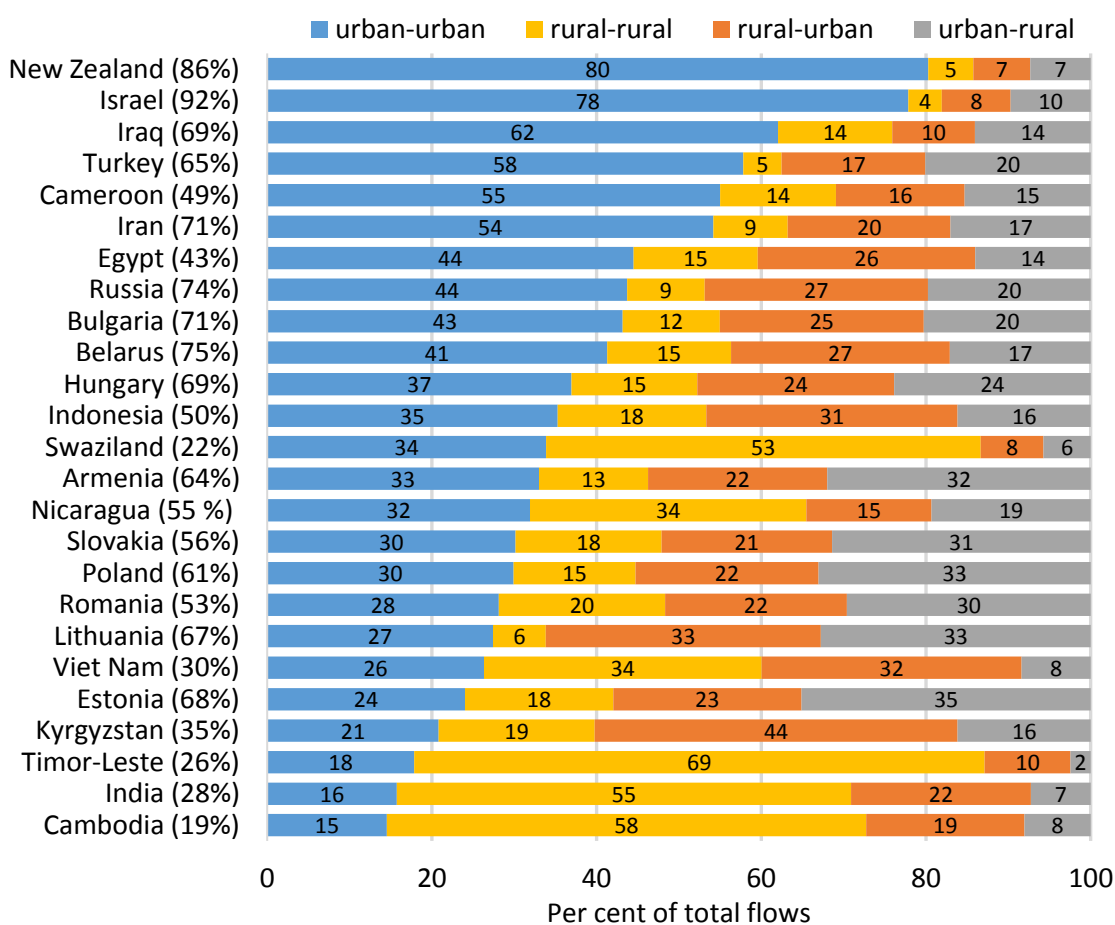

**Figure 12** Shares of migration between rural and urban areas, selected countries

*Note: Level of urbanisation in parentheses*
*Source: Rees et al. (2016)*

To what extent are different educational groups over- or under-represented in these four sets of flows between rural and urban areas? Only 12 countries in the IPUMS dataset available for this study collect data which identifies whether migrants were living in a rural or urban locality prior to migration. Figure 13 depicts the proportion of the flow in each of these countries that is comprised of people who have completed secondary or higher levels of education. Aggregate levels of education vary widely between the sample countries but a number of consistent patterns emerge. Most notable is that in all 12 countries rural-to-urban migrants are on average more educated than those moving from one rural area to another, with the differences being especially pronounced in India, Nicaragua and Cameroon. Migration in the reverse direction, from urban to rural areas, shows a slightly weaker educational mix in some countries (Kyrgyzstan, US, Nicaragua, Cameroon), but includes more well educated people in others (Belarus, Egypt, Ethiopia). In general, however, the two-way flows between urban and rural areas appear closely matched in terms of educational attainment, and this is especially notable in Armenia, India, Nicaragua, Iran and Iraq. Of the 12 countries examined here, only Thailand shows a significantly better-educated profile moving from rural to urban areas. Also apparent from Figure 13 is that it is moves from one urban area to another in which the well-educated are most strongly represented, with Thailand again offering the sole exception. This inter-urban profile is readily explained as a product of the fact that urban residents, on average, have a higher level of education than people living in rural areas, and urban areas in turn offer a range of social, educational and employment opportunities better suited to those with higher levels of schooling.

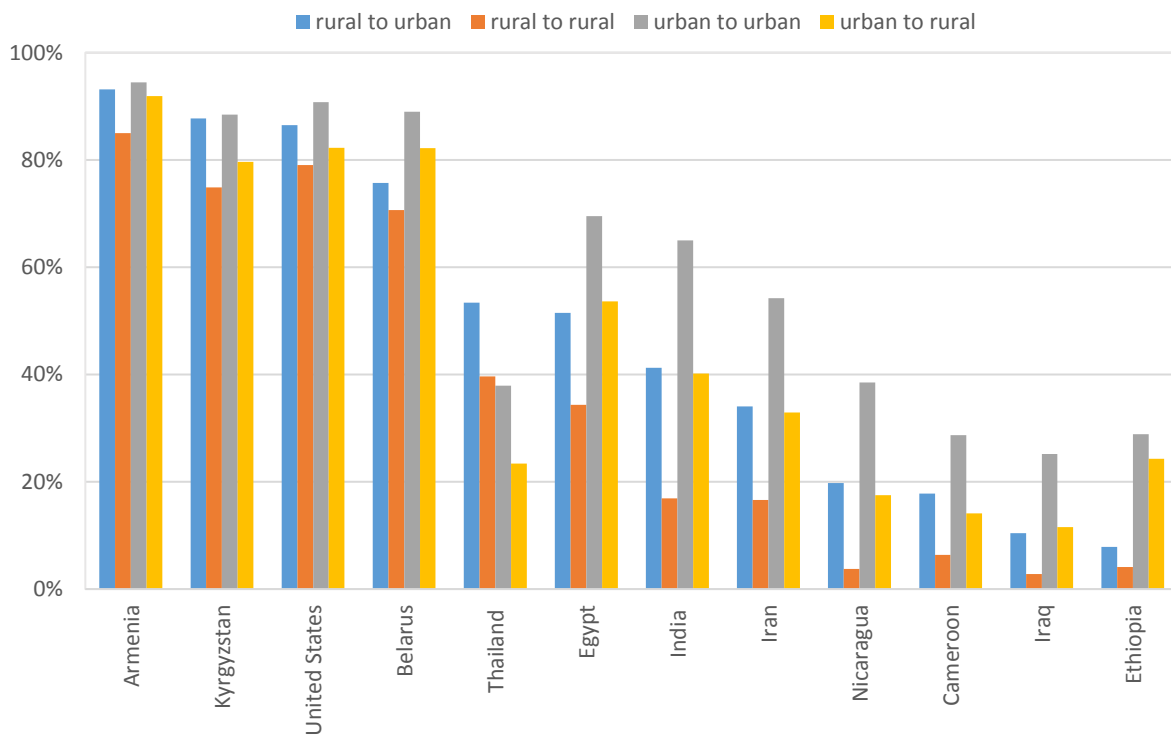

**Figure 13** Percentage of migrants with at least secondary education completed

*Note: migration between major regions*

Although few countries collect rural-urban migration directly, it is possible to cast further light on this form of movement by comparing the educational attainment of urban in-migrants against non-migrants living in urban and rural areas. This is because the urban or rural status of the current place of residence is known for all countries in the IPUMS sample, and we are able to identify those whose migration terminated in an urban area, irrespective of origin. By focusing on migration between major regions, we can largely eliminate from the migrant group those who simply changed residence within the same urban area. Drawing on this information, we compare the educational attainment of urban in-migrants (migrants who have moved to urban areas irrespective of their previous place of residence) against that of rural non-migrants and urban non-migrants. For ease of interpretation mean years of schooling has been used as a measure of educational attainment. This is available for 29 countries in Africa, Asia and Latin America.

Figure 14 plots mean years of schooling for urban in-migrants and urban stayers for each country, with the principal diagonal dividing the plot where mean years of schooling is equal for both groups. Since most observations fall below the diagonal, it is apparent that urban in-migrants tend to have higher levels of education that urban stayers, with mean years of schooling on average 1.11 times higher among the in-migrants. In Mali, Malawi and Uruguay this ratio is just 1.05, whereas educational differences between urban stayers and movers are much more significant in Rwanda (1.43), Indonesia (1.25), Thailand (1.21), Vietnam (1.20) and Cameroon (1.20). Only in Ecuador, Guinea, Haiti and Kenya do urban in-migrants have fewer mean years of schooling than urban stayers. These ratios appear to be unrelated to the overall level of education within each country: no association was found between country-level mean years of schooling and the degree of educational selectivity of urban migrants, measured by the ratio of mean years of schooling of urban in migrants to that of urban stayers.

The positive educational selectivity of urban in-migrants is even more apparent when compared to rural stayers as demonstrated in Figure 15. In all countries, urban migrants are positively selected with regard to education, with mean years of schooling averaged across all countries 1.57 times higher than that of rural stayers. In seven of the 29 countries, mean years of schooling are 1.1 to 1.5 times higher for urban in-migrants than for rural stayers, for 12 countries the ratio sits between 1.5 and 2 and for 10 countries the ratio ranges from 2 to 6.

Unlike the comparison to urban stayers, Figure 16 reveals that the degree of selectivity of urban in-migrants, compared with rural stayers, is strongly conditioned by the overall level of schooling in a country. As mean years of schooling in a country rises, the ratio of schooling among urban in-migrants to that of rural non-migrants falls rapidly. In countries where the average level of education is relatively low, especially in Africa, urban in-migrants report much higher levels of education that than rural stayers. In Guinea, Mali and Senegal, where mean years of schooling is around 2 years, urban in-migrants on average spent four times as many years in school than rural stayers. In Chile and Jamaica, on the other hand, where schooling averages around nine years per person, urban in-migrants have attended school for only 1.6 and 1.2 times as many years as rural stayers. The net effect is that in countries with relatively low levels of education overall, the impact of rural outmigration will be especially pronounced, since it is likely to be the more educated people who will move to urban areas.

This result offers solid confirmation for the proposition advanced by Gould (1982) that as education expands, the educational selectivity of rural to urban migrants decreases. Figure 16 quantifies this effect by tracing a power relationship between the two variables according to which additional years of schooling exert a steadily diminishing effect on the selectivity ratio. The power function in Figure 16 is strongly conditioned by the three African countries with low national average years of schooling, for which the selection ratios are especially high. Even if these three observations are set aside, the negative effect of national advances in education is sustained, with the remaining countries delivering a linear relationship of the form y=-0.15x +2.71 (r-squared=0.55). According to this function, the ratio of years schooling among urban in-migrants to years schooling among rural stayers would fall from 2.41 for countries with an average of two years schooling to 1.36 for those with an average of nine years schooling. That is, each additional year of national schooling would result in a 6 per cent reduction in the selectivity ratio.

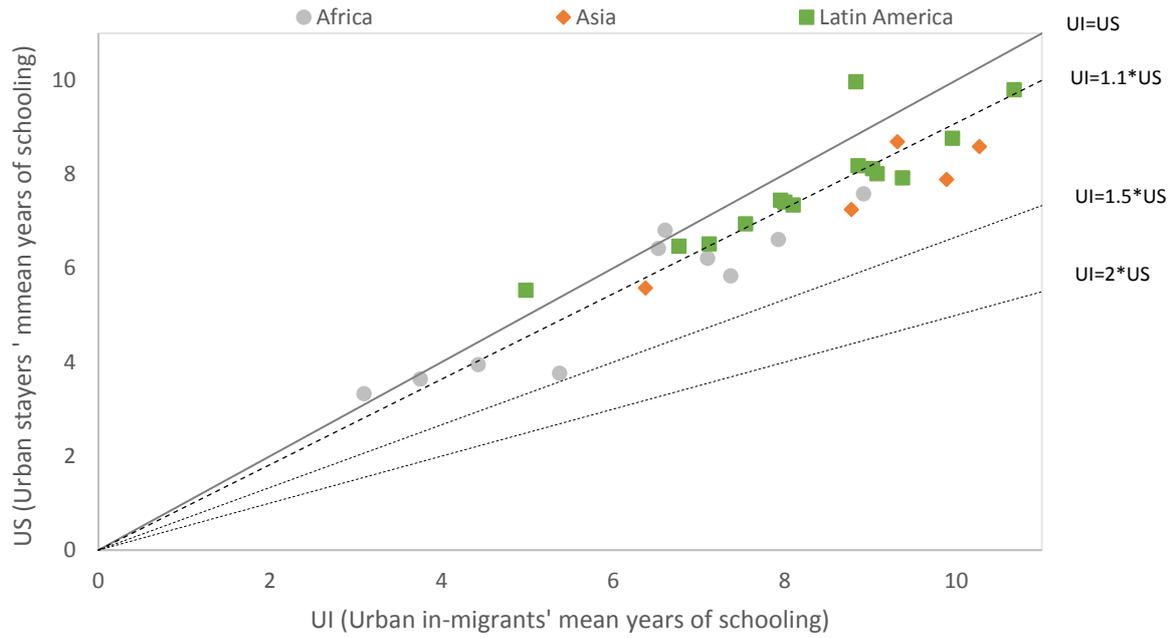

**Figure 14** Mean years of schooling by migration status, individuals aged 15 years and over

*Note: migration between major regions*

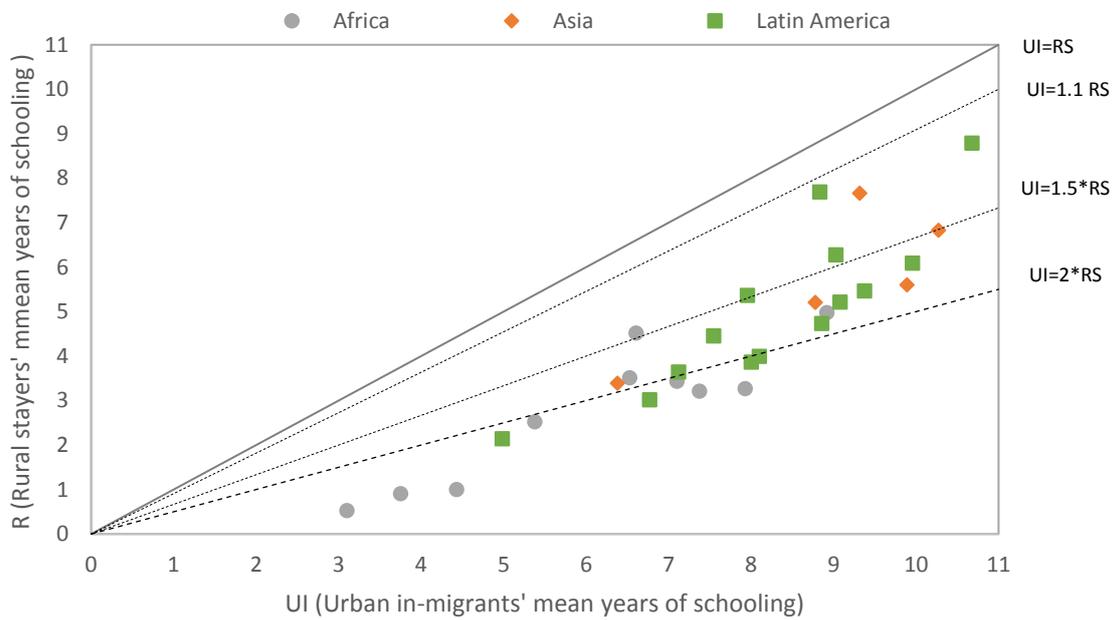

**Figure 15** Mean years of schooling by migration status, individuals aged 15 years and over

*Note: migration between major regions*

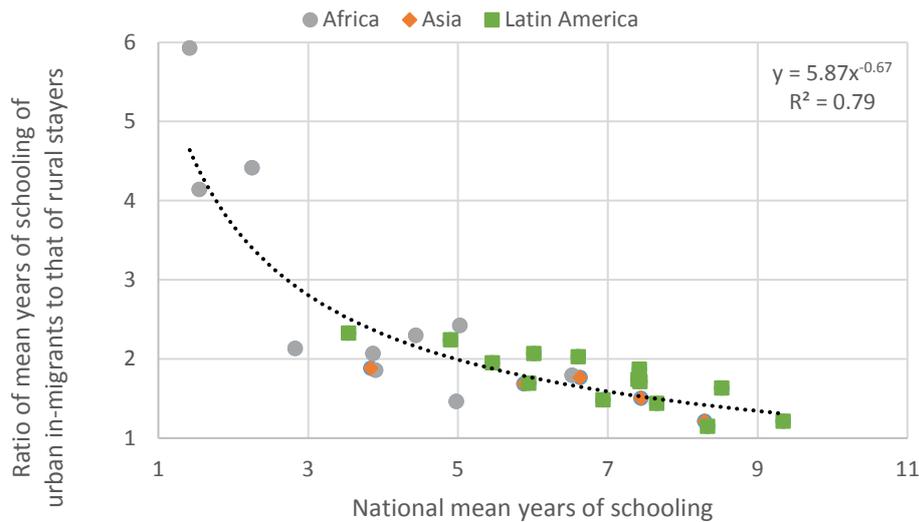

**Figure 16** Mean years of schooling by migration status

*Note: trend line represents a power fit, migration between major regions*

We further test the educational selectivity of urban in-migrants by replicating the analysis for young adults aged 20 to 24. Results are reported in Table A8 of the Appendix and show, as one would expect, that young adults on average are more educated that the general population. Across the 29 countries in our sample, young adults report mean years of schooling 23 per cent higher on average than the general population. Among non-migrants, the difference is larger, with urban stayers showing a 36 per cent increase and rural stayers a 49 per cent increase in mean years of schooling compared to the general population of stayers. Table 3 reports the ratio of schooling of urban in-migrants to that of urban and rural stayers and shows that in all countries urban in-migrants aged 20 to 24 are positively selected in terms of education compared with rural stayers, with an average ratio of 1.75. Results are mixed, however, when urban in-migrants are compared with urban stayers. In 11 of 29 countries young adults moving to urban areas report lower average schooling that urban stayers and, while in 18 countries urban in-migrants are positively selected, the difference compared with urban stayers is modest. These result suggests that urban in-migrants are aspirants. They have higher levels of educational attainment that rural stayers but generally not yet as high as urban residents who have had better access to opportunities.

|  |  | 20-24 years old | | 15 years and over | |
| --- | --- | --- | --- | --- | --- |
|  |  | Ratio of mean years of schooling of urban in-migrants to that of urban stayers | Ratio of mean years of schooling of urban in-migrants to that of rural stayers | Ratio of mean years of schooling of urban in-migrants to that of urban stayers | Ratio of mean years of schooling of urban in-migrants to that of rural stayers |
| Africa | Mali | 1.00 | 2.87 | 1.03 | 4.14 |
|  | Senegal | 1.01 | 3.88 | 1.12 | 4.42 |
|  | South Africa | 1.04 | 1.26 | 1.18 | 1.79 |
|  | Cameroon | 1.07 | 2.11 | 1.20 | 2.43 |
|  | Ghana | 1.12 | 1.86 | 1.26 | 2.30 |
|  | Guinea | 0.64 | 6.43 | 0.93 | 5.93 |
|  | Malawi | 1.00 | 1.62 | 1.02 | 1.86 |
|  | Kenya | 1.09 | 1.43 | 0.97 | 1.46 |
|  | Uganda | 1.01 | 1.68 | 1.14 | 2.07 |
|  | Rwanda | 1.06 | 1.48 | 1.43 | 2.13 |
| Asia | Fiji | 1.01 | 1.11 | 1.07 | 1.22 |
|  | Cambodia | 0.92 | 1.43 | 1.14 | 1.88 |
|  | Indonesia | 1.08 | 1.40 | 1.25 | 1.76 |
|  | Vietnam | 1.09 | 1.30 | 1.20 | 1.50 |
|  | Thailand | 0.99 | 1.28 | 1.21 | 1.69 |
| Latin America | El Salvador | 0.95 | 1.46 | 1.09 | 1.95 |
|  | Costa Rica | 1.02 | 1.18 | 1.11 | 1.44 |
|  | Dominican Republic | 1.03 | 1.26 | 1.07 | 1.48 |
|  | Ecuador | 0.95 | 1.21 | 0.89 | 1.15 |
|  | Argentina | 1.05 | 1.38 | 1.18 | 1.72 |
|  | Bolivia | 0.91 | 1.57 | 1.08 | 2.07 |
|  | Chile | 1.02 | 1.31 | 1.14 | 1.63 |
|  | Columbia | 0.99 | 1.64 | 1.10 | 2.03 |
|  | Haiti | 0.86 | 1.63 | 0.90 | 2.33 |
|  | Jamaica | 1.03 | 1.10 | 1.09 | 1.21 |
|  | Mexico | 1.00 | 1.28 | 1.13 | 1.74 |
|  | Nicaragua | 0.94 | 1.88 | 1.05 | 2.24 |
|  | Paraguay | 0.93 | 1.37 | 1.09 | 1.69 |
|  | Peru | 0.98 | 1.39 | 1.08 | 1.87 |
|  |  |  |  |  |  |
|  | Mean | 0.99 | 1.75 | 1.11 | 2.11 |

**Table 3** Ratio of mean years of schooling, urban in-migrants compared with urban and rural stayers, selected ages

## 8. The effects of duration of residence

Among young adults, migration to urban areas aims, at least for some, to enhance skills and educational qualifications, but such aspirations are not realised overnight. How does migration to urban areas play out over time in terms of migrant educational attainment? This section endeavours to shed light on this issue by analysing educational attainment according to the duration of residence in urban areas of migrants from rural origins. This analysis aims to help establish whether longer durations of residence in urban areas are associated with greater educational attainment. As shown in Appendix Table A1, the requisite data on duration of residence and urban status of previous place of residence are only available for eight countries: India, Ethiopia, Egypt, Cameroon, Belarus, Iraq, Kyrgyzstan and Thailand.

Figure 17 reports the proportion of rural-to-urban migrants with at least secondary education by duration of residence in urban areas. The results show no change in the level of educational attainment among migrants with progressively longer durations of residence in urban areas up to five years. As demonstrated earlier, urban in-migrants already display levels of education similar to (age 20-24) or higher than (age 15+) urban stayers, and only a small proportion of migrants move for educational purposes, so the overall increment is unlikely to be large. Figure 18 uses data on mean years of schooling to provide a finer measure of year-to-year changes in educational attainment for Cameroon and Thailand for all types of migrants (rural-urban, urban-urban, urban-rural and rural-rural). The results confirm that there is no significant change in levels of educational attainment for longer durations of residence for all origins and destinations. It is important to note that Figures 17 and 18 are a synthetic composite in which migrants with different durations of residence arrived in urban areas at different points in time and potentially for different reasons, so the results should be interpreted as indicative rather than conclusive. Ideally, the effect of duration of residence needs to be explored using longitudinal analysis or retrospective survey data, encompassing a lengthier period of residence and extended to consider the effect on subsequent generations, especially the educational attainment of migrants' children as well as the migrants themselves.

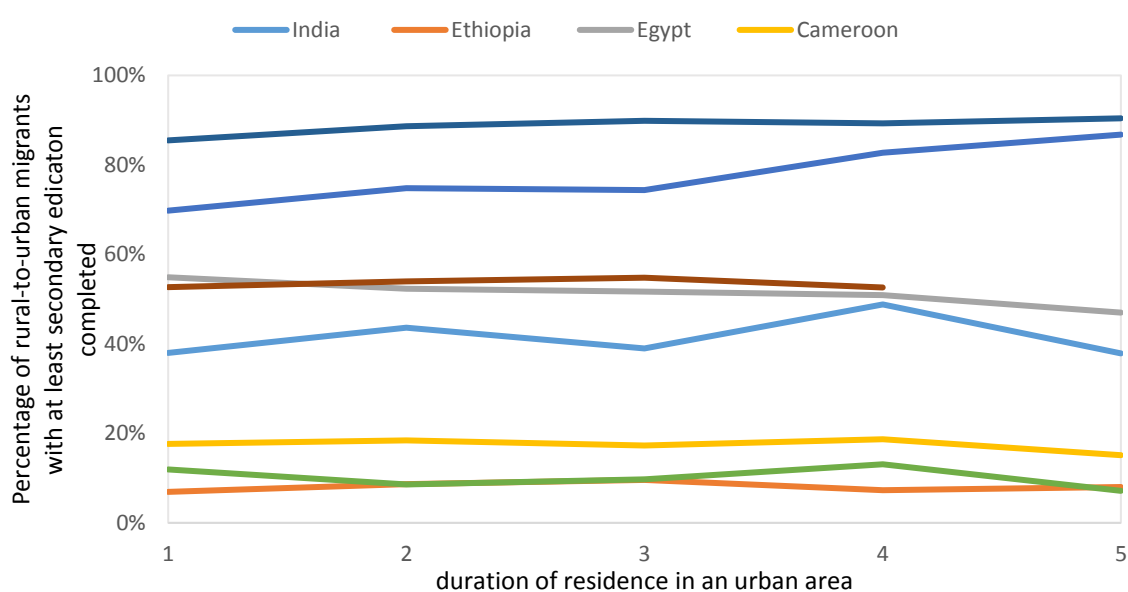

.

**Figure 17** Percentage of rural-to-urban migrants with at least secondary education completed by duration of residence in an urban area

*Note: migration between major regions, Thailand collects duration of residence up to 4 years*

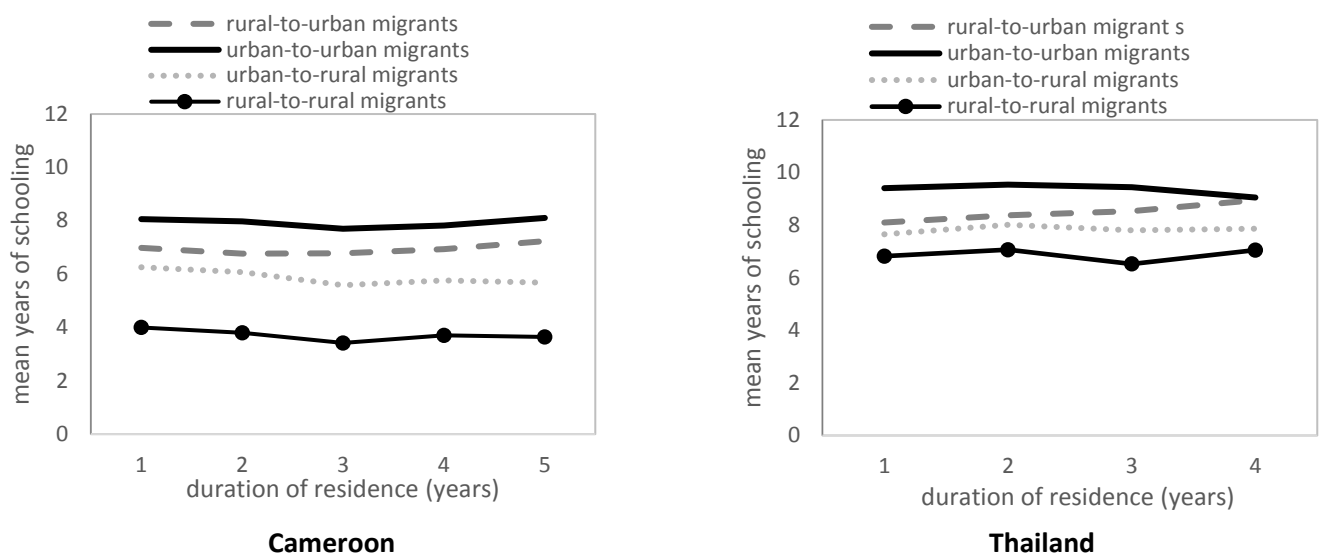

**Figure 18** Mean years of schooling by duration of residence and migrant type

*Note: migration between major regions, Thailand collects duration of residence up to 4 years*

## 9. Migration and the Redistribution of Human Capital

Migration is a key process redistributing human capital and is widely seen as instrumental in the movement of skilled labour from rural to urban areas. Section 7 provided qualified support for this view with rural to urban flows generally including higher proportions of educated people than flows in the reverse direction. However, the pattern was by no means consistent, and data were available for only a small number of countries. As pointed out earlier, very few countries collect data which clearly distinguish moves between rural and urban areas, and reliable comparisons are further compromised by inconsistent definitions of 'rural' and 'urban'. In any event, it is clear that the broad dichotomy between 'rural' and 'urban' areas is overly simplistic as a representation of the complex settlement patterns that characterise societies in the contemporary world.

To circumvent these issues, we adopt an alternative solution based on the ideas developed by Courgeau (1992), elaborated by Rees and Kupiszewski (1999) and implemented into a general model by Rees et al. (2016) which involves using average population density in each region of a country as a proxy for the conventional urban/rural classification. We then analyse the pattern of net migration gains and losses for different educational groups, the aim being to establish whether each is moving, on average, to more or less densely populated regions. For each educational group in a country, we calculate the net migration rate for each region, and set this against the logarithm of the region's population density, with the nature of the relationship summarised through the slope of a simple linear regression. As illustrated for the total population in Figure 19, a positive slope in Haiti signals a general pattern of population losses from sparsely populated areas (at the left end of the x-axis) and corresponding gains in the more densely populated parts of the country (at the right). A negative slope shows the reverse relationship in France, whereby less densely populated regions are gaining, which corresponds to counter-urbanisation. Thus, the sign of the regression parameter indicates whether migration is taking place from predominantly rural to predominantly urban areas (less settled to more settled) and the value of the slope indicates the strength of the movement.

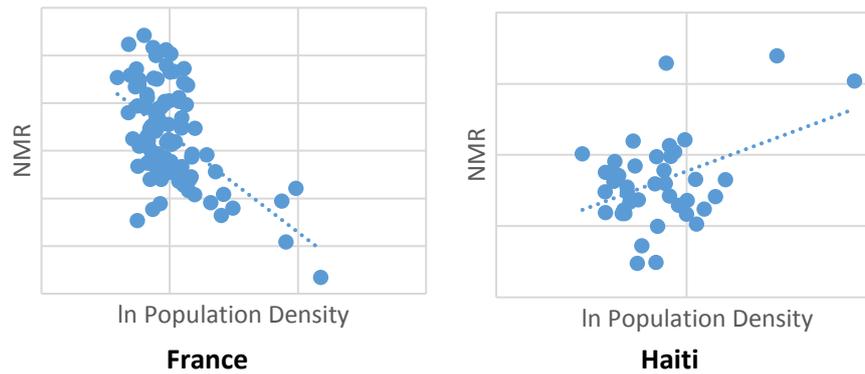

**Figure 19** Relationship between net internal migration rates and the logarithm of population density of regions, selected countries

*Note: NMR is the Net Migration Rate. Each dot correponds to a region within the country.*

Rees et al. (2016) embedded these ideas in a general conceptual model based on observations from 67 countries that depicts how the regression line shifts with the process of development: starting as a shallow positive slope in the early stages of rural to urban migration, steepening as the rate of urbanisation gathers pace, falling back in the latter stages of national development, then oscillating between shallow positive and negative slopes. While the model appears well suited to capturing contemporary patterns of population redistribution at the aggregate level, the regression approach can also be readily applied to examine the redistribution of individual educational groups.

To establish the context, we first explore the relationship between net migration rates and population density for 48 countries focusing on the aggregate population. The results are shown separately for each region in the four panels of Figure 20 and reveal significant variations between and within regions in the redistributive effect of migration. Steep, positive slopes indicate high levels of migration from low density regions (rural settlements) to high density regions (urban settlements), leading to the largest net migration gains in the most densely populated regions. This pattern is characteristic of countries with low but accelerating levels of urbanisation, such as Uganda, Kyrgyzstan and Haiti, with overall urbanisation rates below regional averages. Positive but moderate slopes also point to net movements from low to high density regions but at a slower rate and are found in more urbanised countries including Cameroon, Morocco, China, Thailand, Honduras and Peru. A third group of countries display relatively flat slopes, either positive or negative, indicating that migration has a more limited redistributive effect on the settlement of populations. This group features some countries at early stages of urbanisation, such as India and Mali, where migration from low to high density regions remains limited but is expected to increase as the process of urbanisation proceeds. Flat slopes characterise mainly highly urbanised countries where migration flows between urban and rural areas are relatively balanced with little population redistribution. Some of these countries, such as Iraq, Brazil, Mexico and Switzerland display moderate positive slopes, while in others the modest negative slopes indicate that the direction of migration has reversed in a way that net gains favour less densely populated areas. This pattern of redistribution is found in Argentina, Uruguay, Spain, Portugal but also Iran, Ghana and Indonesia. In France and the United States negative slopes are steeper, indicating that the largest net migration losses occur from the most densely population regions, which signals a process of counter-urbanisation where there is a strong preference for low density living. These results accord closely with the theoretical expectations based on the model proposed by Rees et al (2016) which sees a systematic relationship between migration and population density as national development proceeds.

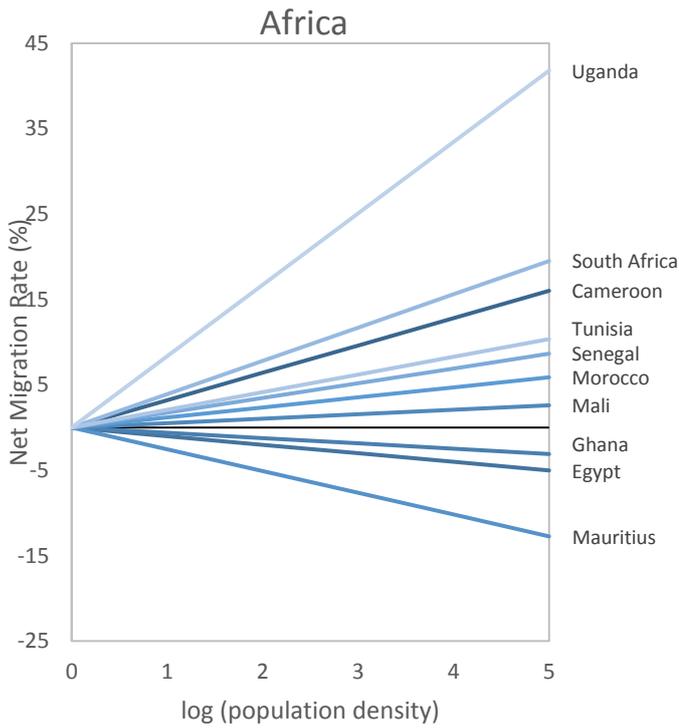
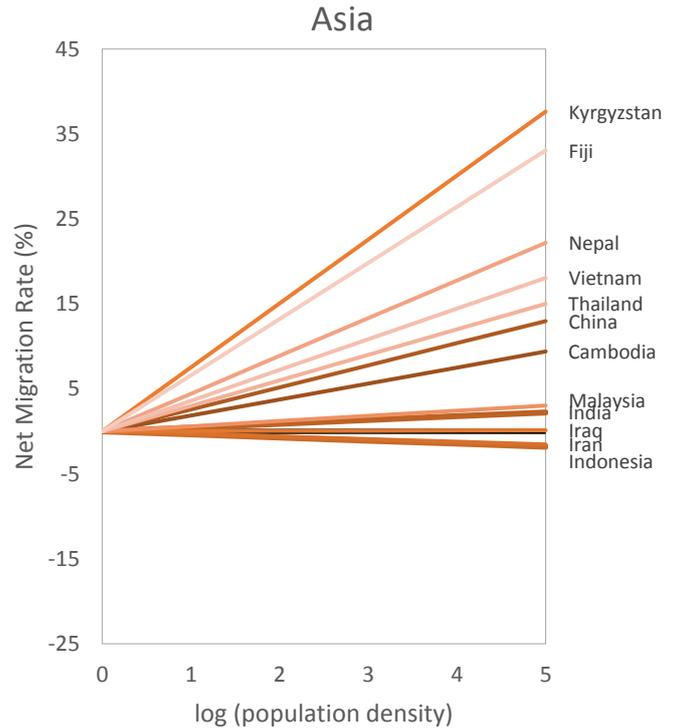
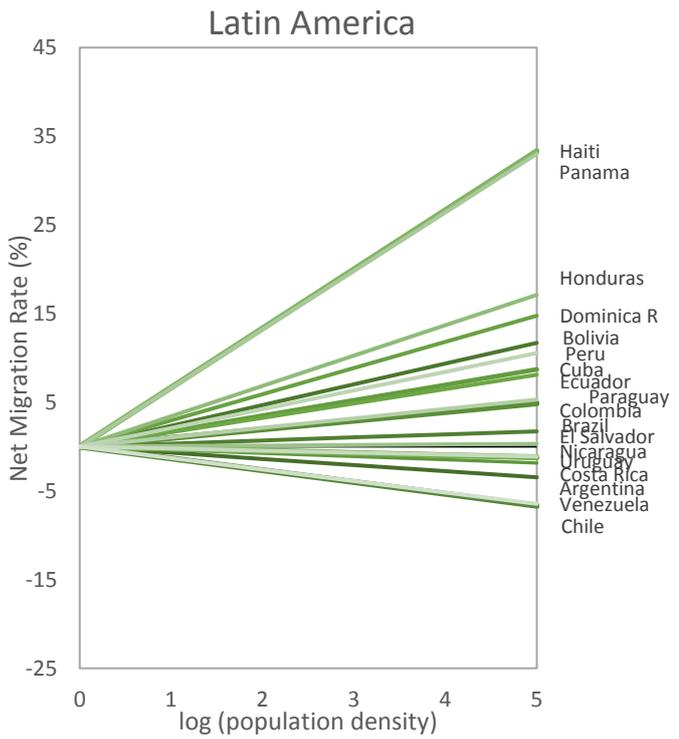
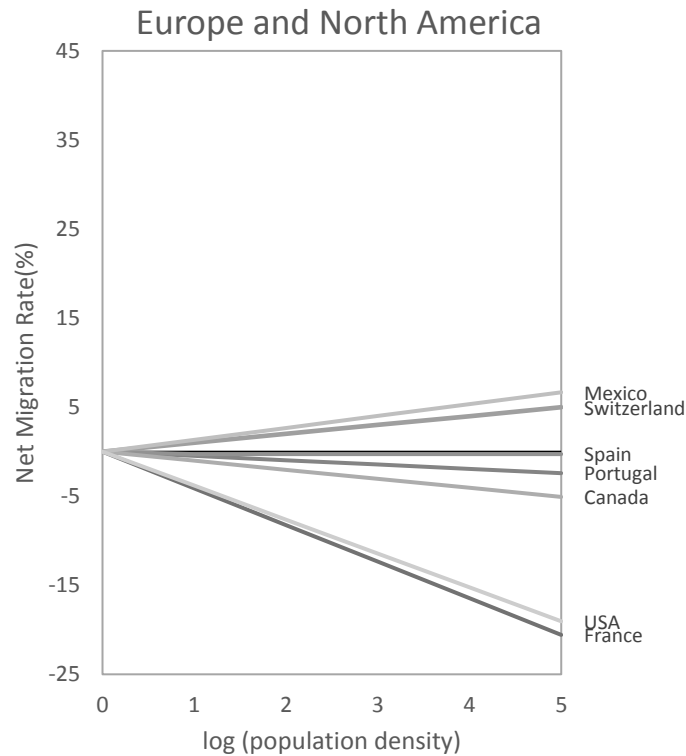

**Figure 20** Fitted slopes capturing the relationship between net migration rates and population density, selected countries by region

Having established the extent and direction of population redistribution, we now turn attention to variations in the impact of migration by educational attainment by replicating the above analysis for individual segments of the population classified by educational attainment. Figure 21 reports results for a sub-sample of six countries with diverse slopes and shows that the direction and strength of population redistribution varies significantly between countries, but that greater consistency is found across categories of educational attainment in some countries, though not in others. Care is needed in interpretation because the vertical axis varies between graphs. Figure 22 draws together the slope parameters for the six countries into a single graph which more readily highlights the underlying patterns.

Variations between educational categories in the extent of redistribution are pronounced in the two countries where migration has a very strong redistributive effect, either from low density to high density regions (Uganda) or from high density to low density regions (United States). In Uganda, the slope of net migration rates against the logarithm of population density is steep and positive and forms a distinctive inverted U-shape as educational attainment increases. This means that the large net migration gains observed in the most densely populated regions are especially pronounced for populations with primary and secondary education and are more moderate for populations with less than primary education or with tertiary education. A similar profile is apparent in the United States but with precisely the reverse effect. The steep negative slope indicates a strong preference for low density living but in this case the U-shaped relationship to education means it is those with primary and secondary education who are driving net population gains in the least densely populated regions. At either end of the educational spectrum, slopes are relatively flat, especially for individuals with less than primary education, indicating that migration of these groups in the US has a very limited redistributive effect on the settlement of these populations.

In Haiti and Senegal, both of which are experiencing a net movement from lower to higher density regions, it is again those with primary and secondary education who record the highest rates of redistribution. Among those at the ends of the educational spectrum – those with tertiary education and those with less than primary – the redistributive effect of migration on population distribution is much less pronounced. The direction and composition of the movement therefore echoes that in Uganda, although the extent of the redistribution is considerably weaker, and the redistribution of tertiary qualified individuals is especially low. In Haiti and Senegal, as in other developing countries, this almost certainly reflects the limited opportunities for highly skilled people outside the major cities.

Argentina and China present profiles quite different from the former countries. In Argentina, the slope is relatively flat and negative at all levels of educational attainment, but it is notable that those with tertiary education appear to be moving away from more densely populated regions at a slightly faster pace than groups at other levels of educational attainment. In China, on the other hand, it is the tertiary-educated who exhibit the fastest rate of redistribution to more densely settled areas. Indeed, there is a steady increase in the slope with rising levels of educational attainment, but migration has almost no redistributive effect on the settlement of individuals with less than primary education.

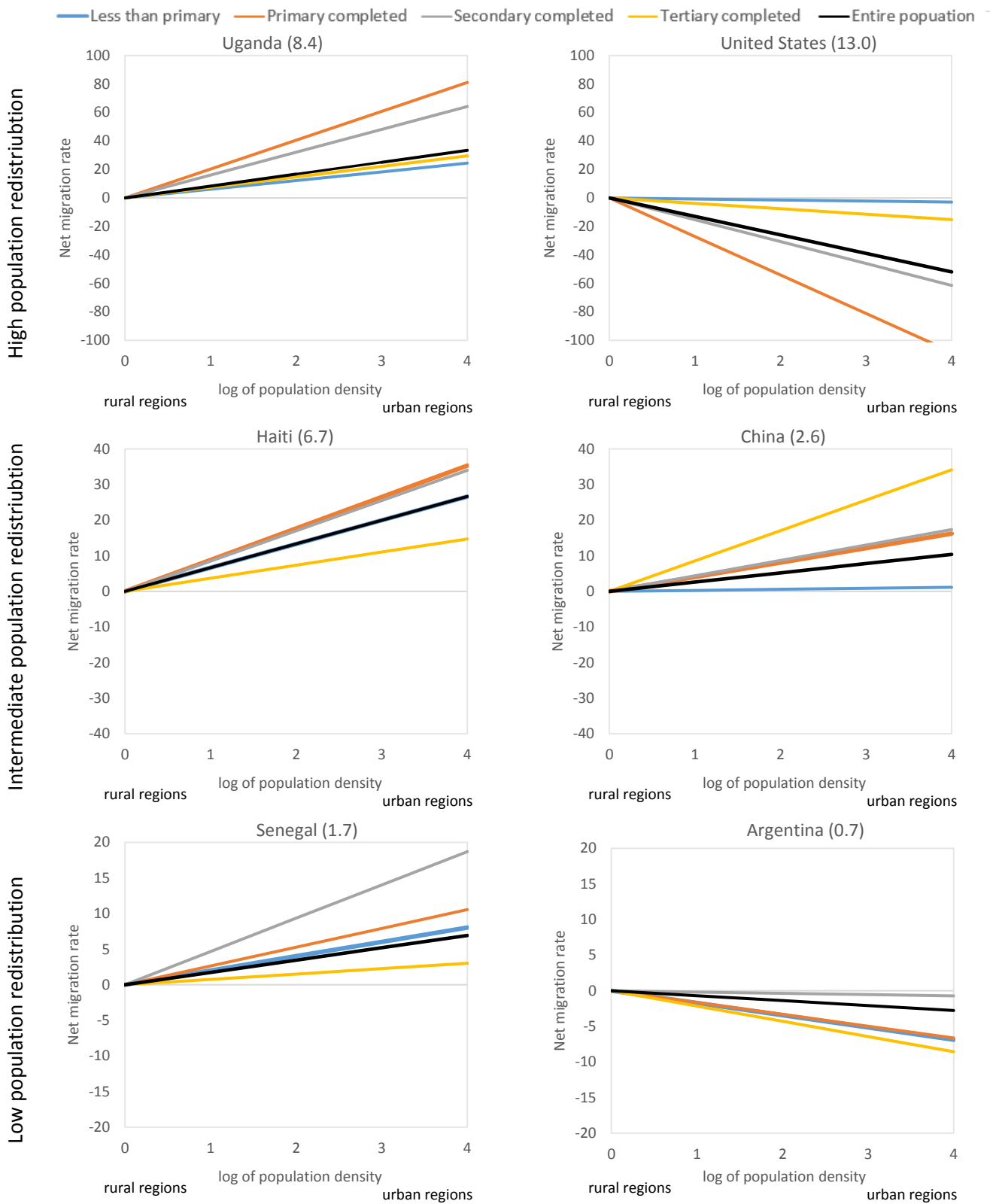

**Figure 21** Regression coefficients of net migration rates against the population density of individual regions by educational attainment, selected countries

*Note: Results for migration between major regions. Regressions are population weighted. Correlation coefficients for the total population are indicated in brackets. Note the difference between the six graphs in the scale on the vertical axis.*

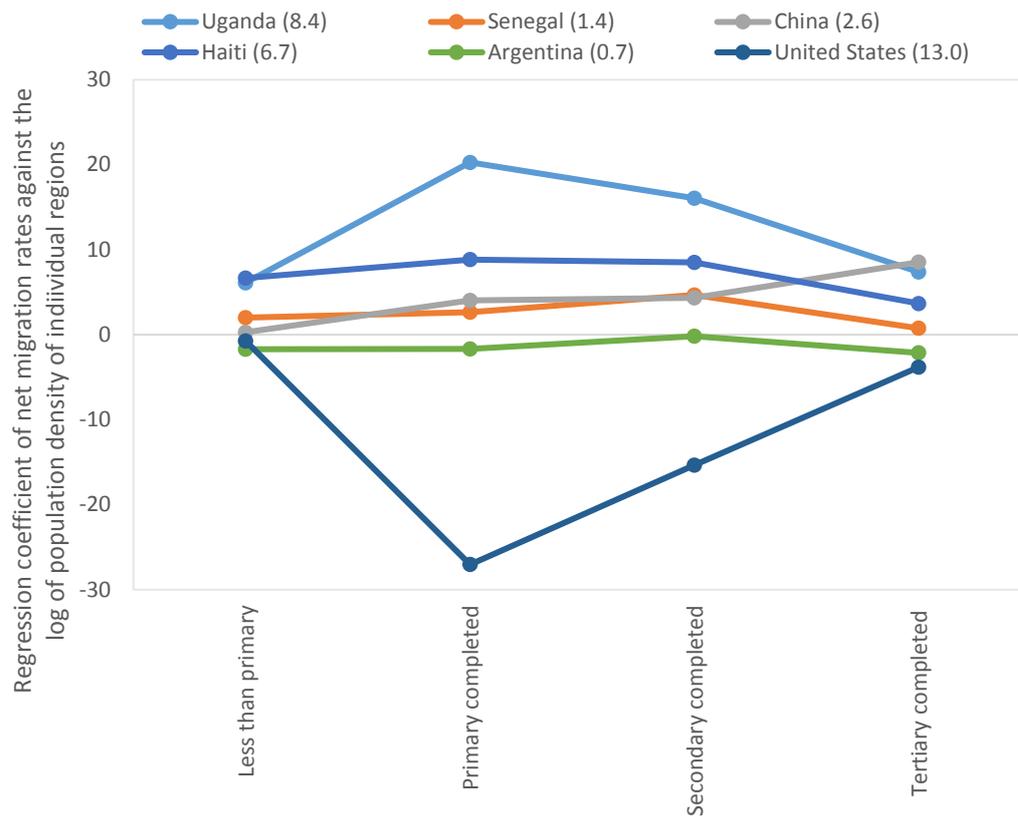

**Figure 22** Regression coefficients of net migration rates against the population density of individual regions by educational attainment, selected countries

*Note: regression coefficients for the total population are in brackets, regression coefficients by educational attainment are for the population aged 15 and over. Results for the United States are for migration between states (counties on Figure 20)*

# 10. Summary and Conclusion

Education is a fundamental building block in the process of national development, and migration is instrumental both in the acquisition of this human capital and in its redistribution to centres of economic activity. A burgeoning literature has provided useful insights into some of the links between these two phenomena, but comprehensive understanding has been severely constrained by limitations in the available data, as well as in the scope of research. For the latter, one notable bias in recent studies has been a primary focus on international migration to the neglect of movements within individual countries. As a result, the current state of understanding of the reciprocal links between migration and education is best described as partial, fragmented and unsystematic. This report aimed to redress this imbalance and enhance understanding by applying a recently developed range of analytical techniques to a newly established repository of Census data on internal migration in countries around the globe.

We focused on three key aspects of migration: its overall intensity, spatial patterning, and redistribution, differentiating educational groups based variously on level of attainment and years of schooling, and distinguishing countries according to regional (continental) groupings and levels of HDI. For each facet of migration, we first sought to establish the global context by reviewing the scale and diversity of migration behaviour in the population at large, drawing on a large sample of countries. Migration is commonly triggered by educational transitions at young adult ages, and the spread of education over time is reflected in marked differences in levels of educational attainment across the age profile. We therefore gave particular attention to the role of age in shaping the intensity of migration, and to the age composition of migrants.

With respect to migration intensity, we found a strong, positive association between level of education and the propensity to move. The tertiary-educated display the highest rates of movement, followed by those who completed secondary education, those with only primary schooling, and those who failed to complete primary. Expressed as an educational gradient, completion of primary education effectively doubles the probability of migration, secondary education triples it, and tertiary education quadruples it, compared to those with no formal education. This multiplier effect is greatest for movement over longer distances, that is between major regions such as states and provinces, and a little less pronounced for shorter distance moves. With very few exceptions it is pervasive across all continents and is especially strong in Asia and Africa.

The link between migration and education is found for groupings of countries at all levels of HDI, but the effect of education on mobility appears to diminish with progress through the educational transition, as suggested by Gould (1982). The evidence assembled in this report reveals that this involves a systematic, graded change as countries develop. Primary and secondary education exert the most pronounced effect in raising levels of migration at the lower end of the development ladder, but these differentials steadily diminish, so that tertiary education remains as the key factor in educational selectivity among economically advanced countries. The evidence assembled here also showed that differentials in migration intensity between groups at different levels of educational attainment are remarkably stable across the age spectrum. Higher rates of movement among secondary and tertiary educated people are found at all ages, not simply among young adults, and as the differentials diminish with development, this occurs throughout the age profile.

This transition in migration intensities is echoed in the composition of migration flows, and amplified by population age structures, such that marked differences are found in the educational composition of migrants in countries at different stages of development. At the lower end of the development spectrum, high proportions with less than secondary education generate migration flows that are heavily weighted towards lower levels of educational achievement, despite above average migration intensities among these groups. Conversely, the more advanced educational profiles characteristic of populations in more developed countries combine with very high migration intensities to generate migration streams with much larger shares of secondary and tertiary educated people.

Age is a key factor shaping all forms of migration, but is especially significant to education because the transition into post-secondary study, and subsequently into the labour force, often involves a residential relocation. Moreover, these transitions commonly occur during the young adult years, where migration is at its peak. Indeed, education represents one of the four major transitions to adulthood that underpin the peak in migration that is found in the early to mid-twenties. Migration often occurs for a combination of motives and few countries identify the reasons for migration at the Census but, among those that do, education generally emerges as a minor reason. Its significance varies between countries but appears to increase with national development, although our sample is too small to provide a definitive conclusion. What is common among all countries is that moves for education are highly age-graded, being concentrated into a relatively narrow and symmetrical age band, and that they peak at an earlier age than moves triggered by other life course transitions, such as employment or marriage. Despite these commonalities, we found subtle variations between countries in the age structure of education-related migration that reflect their unique social, cultural and economic contexts. China, in particular, stood out with an early and highly pronounced peak for education-related migration. Sex differences are also striking and in many countries education appears to be predominantly a male rationale for migration: this is most pronounced in India and the Middle-East.

Migrants are generally thought to be selective of the population at the origin, but less well qualified than those at their destination. Our results provide partial support for this view. There is clear evidence that the composition of migration flows differs according to the origin and destination of the migration stream. Few countries collect the data needed to rigorously clearly establish the composition of moves between country and city, but rural to urban flows include higher proportions of educated people than flows from one rural area to another and the educational profile is strongest in flows between urban areas. Positive educational selectivity is clearly apparent when comparing urban in-migrants to rural stayers. Evidence assembled from a sample of 29 countries shows that urban in-migrants on average have 1.57 times as many years of schooling. Using these data, urban in-migrants also emerge as having higher levels of education that urban stayers, with mean years of schooling on average 1.11 times greater. These results are shaped by differences in age composition, since migrants are younger than those at either the origin or destination, and the positive selection compared with urban stayers all but disappears if attention is confined to young adults. However, the results also indicate that the degree of selectivity of urban in-migrants, compared with rural stayers, is strongly conditioned by the overall level of schooling in a country. As mean years of schooling in a country rises, the ratio of schooling among urban in-migrants to that of rural non-migrants falls rapidly. In countries with relatively low levels of education overall, the impact of rural outmigration will therefore be especially pronounced, since it is the better educated who will move to urban areas. Since some migrants move to the cities to further their education, we also expected that the differential between urban stayers and urban in-migrants would increase as the latter enhanced their education with rising duration of residence. From the analysis presented here, however, we were unable to detect any duration of residence effect.

Migration is seen as a key process in the spatial redistribution of human capital, and our results reveal that this plays out in varying ways across countries at differing levels of national development. Data on movement between rural and urban areas are sparse and difficult to interpret, so we examined the direction and strength of net migration gains (and losses) by educational group according to the overall population density of a country's regions. Theoretical models (Rees et al. 2016) hold that movement from low to high density regions first accelerates then slows as the process of national development proceeds, and capture this in the gradient of a simple linear regression. Our results shows that the *direction* of redistribution is generally consistent for all educational groups, denoting either gains or losses in more densely populated areas, and *vice versa*. Moreover, the strength of the redistribution between regions, as denoted by the *gradient* of the regression, accords with expectations from the theoretical model. We also found, that the *gradient* of the regression varies systematically by educational attainment, and appears to be conditioned by a country's level of development. In countries at comparatively early stages of development, such as Uganda, Haiti and Senegal, the pattern of redistribution is from low to high density regions, at varying levels of strength, but in each case it is those with primary and secondary education who register the strongest redistributive effects. The same is apparent in China, except that in that case it is the tertiary educated who are most strongly represented in the redistribution to high density areas. In countries where the net direction of movement is reversed, the patterns are more mixed, but in the USA it is notable that it is those with moderate levels of educational achievement who are most prominent in the process of counter-urbanisation. At advanced stages of development, the tertiary educated appear to be most strongly attracted to, and retained, in the cities. High overall levels of migration intensity among this group therefore appear to be taken up in reciprocal exchanges between areas, rather than in spatial redistribution, resulting in a modest degree of migration effectiveness.

Despite recent advances, the lack of suitably detailed and consistent data on internal migration across countries remains a serious impediment to understanding the links between mobility and education at a global scale. Fundamental to this constraint are differences in the spatial scale to which migration data refer, the so-called Modifiable Areal Unit Problem, which confounds robust comparisons between countries. Equally problematic is the failure of most censuses to distinguish the types of settlement in which migrants were living before they moved, since this precludes a clear identification of key types of move, for example from rural to urban areas. Significant shifts in data collection practice by national statistical agencies are needed to fully address these limitations.

Notwithstanding the advances reported in this paper, further work is needed to better understand and codify the way in which the shape of the mobility profile across educational groups alters as development proceeds. As an increasing number of countries progressively make data from the 2010 census round publicly available, the analysis presented in this report can be extended to a larger number of countries and to multiple observation periods to assess changes in the links between migration and education over time. The opportunity here is to provide a more nuanced understanding of the propositions outlined by Gould (1982) and elaborated here. Ideally we would wish to see the development of a comprehensive theoretical framework that traced temporal evolution in the educational composition of migration flows and in the redistribution of human capital.  Intermediate steps in this process could usefully examine the effects of duration of residence in further detail. Attention is also needed to regional anomalies such as the limited educational selectivity of migrants in Latin America, which might perhaps be due to the high level of urban primacy in the region. The distinctive age profile by reasons for move discussed in this report also invites extension to a wider range of countries to establish whether the observed consistencies hold in different national settings. Notwithstanding the progress made to date, the agenda for research on the links between migration and education remains a rich one.

# Appendix
**Table A1** Migration data

| | Country | Census year | Administrative units (regions) over which migration is measured | | Duration of residence | Urban status (previous residence) | Reasons for moving |
|---|---|---|---|---|---|---|---|
| | | | Minor | Major | | | |
| Africa (n=15) | Botswana | 2011 | | District (n=17) | | | |
| | Cameroon | 2005 | Department (n=58) | Regions (n=10) | X | X | |
| | Egypt | 2006 | | Governorate (n=27) | X | X | X |
| | Ethiopia | 2007 | Wereda (n=529) | | X | X | |
| | Ghana | 2000 | District (n=110) | Region (n=10) | | | |
| | Guinea | 1996 | | Prefecture (n=34) | X | | |
| | Kenya | 2009 | District (n=262) | | X | | |
| | Malawi | 2008 | | District (n=28) | X | | |
| | Mali | 2009 | Circle (n=57) | Region (n=10) | X | | |
| | Morocco | 2004 | Prefecture (n=75) | Region (n=12) | X | | |
| | Mozambique | 2007 | District (n=128) | Province (n=11) | | | |
| | Senegal | 2002 | Department (n=45) | Region (n=14) | | | |
| | South Africa | 2001 | | Province (n=9) | | | |
| | Rwanda | 2002 | District (n=106) | | X | | |
| | Uganda | 2002 | District (n=112) | | X | | |
| Asia (n=14) | Armenia | 2011 | | Province (n=11) | X | X | |
| | Cambodia | 2008 | | Province (n=25) | X | | X |
| | China | 2000 | | Province (n=31) | | | X |
| | Fiji | 2005 | | Province (n=15) | | | |
| | India | 1999 | District (n=683) | State (n=29) | X | X | X |
| | Indonesia | 2005 | Municipality (n=180) | Province (n=34) | X | | X |
| | Iran | 2006 | | Province (n=31) | | X | X |
| | Iraq | 1997 | | Governate (n=19) | X | X | X |
| | Kyrgyzstan | 1999 | | District (n=9) | X | X | |
| | Malaysia | 2000 | District (n=136) | State (n=16) | | | |
| | Nepal | 2001 | District (n=75) | Province (n=7) | | | |
| | Philippines | 2000 | | Province (n=82) | | | |
| | Thailand | 2000 | | Province (n=77) | X | X | X |
| | Vietnam | 2009 | District (n=663) | Province (n=58) | X | | |
| Europe (n=8) | Belarus | 1999 | | Region (n=7) | X | X | |
| | France | 2006 | Department (n=101) | Region (n=26) | | | |
| | Greece | 2001 | Municipality (n=326) | Department (n=8) | | | |
| | Portugal | 2011 | Municipality (n=308) | District (n=20) | | | |
| | Romania | 2002 | | County (n=42) | | | |
| | Slovenia | 2002 | Municipality (n=212) | Region (n=12) | | | |
| | Spain | 2011 | Municipality (n=366) | Province (n=52) | | | |
| | Switzerland | 2000 | Municipality (n=2,294) | Canton (n=26) | | | |
| Latin America (n=17) | Argentina | 2001 | Department (n=324) | Province (n=24) | | | |
| | Bolivia | 2001 | Province (n=112) | Department (n=9) | | | |
| | Brazil | 2010 | Municipality (n=1,540) | State (n=27) | X | | |
| | Chile | 2002 | Municipality (n=178) | Province (n=54) | | | |
| | Colombia | 2005 | Municipality (n=1,104) | Department (n=33) | | | X |
| | Costa Rica | 2011 | Canton (n=81) | Province (n=7) | | | |
| | Cuba | 2002 | Municipality (n=168) | Province (n=16) | X | | |
| | Dominican Republic | 2010 | Municipality (n=154) | Province (n=32) | | | |
| | Ecuador | 2010 | | Province (n=24) | | | |
| | El Salvador | 2007 | Canton (n=236) | Department (n=14) | X | | |
| | Haiti | 2003 | Arrondissement (n=42) | Department (n=10) | X | | |
| | Jamaica | 2001 | | Parish (n=14) | X | | |
| | Nicaragua | 2005 | Municipality (n=152) | Department (n=17) | | X | |
| | Paraguay | 2002 | District (n=224) | Department (n=18) | | | |
| | Peru | 2007 | Province (n=195) | Region (n=25) | | | |
| | Uruguay | 2011 | Locality | Department (n=19) | X | | |
| | Trinidad and Tobago | 2000 | | Region (n=14) | | | |
| North America (n=3) | Canada | 2001 | Census District (n=280) | Province (n=13) | | | |
| | United States | 2000 | County (n=3142) | State (n=51) | | X | |
| | Mexico | 2010 | Municipality (n=2456) | State (n=32) | | | X (2000) |

**Table A2** Human development index, urbanisation rate and mean years of schooling

| region | Country | HDI* | Level of development* | Urbanisation Rate** | Mean years of schooling ***(population aged 25 and over) |
|---|---|---|---|---|---|
| Africa (n=15) | Botswana | 0.698 | medium | 57.4 | 7.4 |
| | Cameroon | 0.518 | low | 53.8 | 5.3 |
| | Egypt | 0.691 | medium | 43.1 | 6.4 |
| | Ethiopia | 0.448 | low | 19.0 | 2.1 |
| | Ghana | 0.579 | medium | 53.4 | 7.0 |
| | Guinea | 0.414 | low | 36.7 | 1.6 |
| | Kenya | 0.555 | medium | 25.2 | 5.5 |
| | Malawi | 0.476 | low | 16.1 | - |
| | Mali | 0.442 | low | 39.1 | 2.0 |
| | Morocco | 0.647 | medium | 59.7 | - |
| | Mozambique | 0.418 | low | 31.9 | 2.4 |
| | Senegal | 0.494 | low | 43.4 | 2.8 |
| | South Africa | 0.666 | medium | 64.3 | 10.3 |
| | Rwanda | 0.498 | low | 27.8 | 3.8 |
| | Uganda | 0.493 | low | 15.8 | 5.1 |
| Asia (n=14) | Armenia | 0.743 | high | 62.8 | 11.7 |
| | China | 0.738 | high | 54.4 | 10.0 |
| | Cambodia | 0.563 | medium | 20.5 | 3.5 |
| | Fiji | 0.736 | high | 53.4 | 9.3 |
| | Indonesia | 0.689 | medium | 53.0 | 7.9 |
| | India | 0.624 | medium | 32.4 | 5.4 |
| | Iran | 0.774 | high | 72.9 | - |
| | Iraq | 0.649 | medium | 69.4 | 7.8 |
| | Kyrgyzstan | 0.664 | medium | 35.6 | 10.9 |
| | Malaysia | 0.789 | high | 74.0 | 10.1 |
| | Nepal | 0.558 | medium | 18.2 | 3.3 |
| | Philippines | 0.682 | medium | 44.5 | 9.1 |
| | Thailand | 0.740 | high | 49.2 | 8.3 |
| | Vietnam | 0.683 | medium | 33.0 | 7.8 |
| Europe (n=8) | Belarus | 0.796 | high | 76.3 | 12.3 |
| | France | 0.897 | very high | 79.3 | 11.3 |
| | Greece | 0.866 | very high | 77.7 | 10.7 |
| | Portugal | 0.843 | very high | 62.9 | 8.9 |
| | Romania | 0.802 | very high | 54.4 | 11.0 |
| | Slovenia | 0.890 | very high | 49.7 | 12.4 |
| | Spain | 0.884 | very high | 79.4 | 9.9 |
| | Switzerland | 0.939 | very high | 73.8 | 13.7 |
| Latin America (n=17) | Argentina | 0.827 | very high | 91.6 | 9.8 |
| | Bolivia | 0.674 | medium | 68.1 | 8.3 |
| | Brazil | 0.754 | high | 85.4 | 7.4 |
| | Chile | 0.847 | very high | 89.4 | 10.0 |
| | Colombia | 0.727 | high | 76.2 | 8.1 |
| | Costa Rica | 0.776 | high | 75.9 | 8.6 |
| | Cuba | 0.775 | high | 77.0 | 11.4 |
| | Dominican Republic | 0.722 | high | 78.1 | 7.8 |
| | Ecuador | 0.739 | high | 63.5 | 8.7 |
| | El Salvador | 0.680 | medium | 66.3 | 6.5 |
| | Haiti | 0.493 | low | 57.4 | - |
| | Jamaica | 0.730 | high | 54.6 | 9.1 |
| | Nicaragua | 0.645 | medium | 58.5 | - |
| | Paraguay | 0.693 | medium | 59.4 | 8.4 |
| | Peru | 0.740 | high | 78.3 | 9.1 |
| | Uruguay | 0.795 | high | 95.2 | 8.7 |
| | Trinidad and Tobago | 0.780 | high | 91.5 | 10.7 |
| North America (n=3) | Canada | 0.920 | very high | 81.6 | - |
| | United States | 0.920 | very high | 81.4 | 13.5 |
| | Mexico | 0.762 | high | 79.0 | 8.6 |

*United Nations, 2015 ** United Nations, 2014 *** UNESCO, 2016*

**Table A3** Sample size by educational attainment

| Region | Country | Total population 15 years and above | Less than primary | Primary completed | Secondary completed | University completed |
|---|---|---|---|---|---|---|
| Africa (n=15) | Botswana | 119,374 | 25,316 | 61,711 | 24,503 | 7,123 |
| | Cameroon | 974,975 | 345,751 | 501,986 | 87,703 | 22,402 |
| | Egypt | 4,958,308 | 2,040,005 | 746,510 | 1,622,118 | 545,131 |
| | Ethiopia | 1,094,113 | 953,128 | 111,981 | 26,310 | 2,694 |
| | Ghana | 1,108,180 | 551,852 | 405,076 | 137,062 | 14,190 |
| | Guinea | 365,259 | 303,701 | 37,612 | 8,456 | 3,009 |
| | Kenya | 2,154,727 | 688,271 | 984,680 | 449,155 | 32,621 |
| | Malawi | 716,794 | 457,073 | 164,545 | 59,797 | 2,761 |
| | Mali | 677,927 | 554,649 | 97,053 | 16,526 | 9,699 |
| | Morocco | 1,332,714 | 950,459 | 269,366 | 87,055 | 25,834 |
| | Mozambique | 1,071,172 | 869,305 | 156,361 | 19,990 | 2,778 |
| | Senegal | 564,518 | 418,902 | 117,302 | 23,315 | 4,999 |
| | South Africa | 2,470,250 | 661,067 | 1,182,240 | 566,655 | 60,288 |
| | Rwanda | 433,271 | 318,389 | 104,749 | 9,177 | 956 |
| | Uganda | 1,254,422 | 684,665 | 494,672 | 66,080 | 9,005 |
| Asia (n=14) | Armenia | 222,085 | 6,755 | 21,121 | 144,696 | 49,513 |
| | China | 9,025,319 | 1,432,433 | 6,108,871 | 1,372,345 | 111,670 |
| | Cambodia | 876,835 | 489,419 | 327,823 | 47,050 | 12,543 |
| | Fiji | 58,184 | 4,148 | 35,919 | 16,863 | 1,249 |
| | Indonesia | 757,593 | 183,033 | 383,746 | 169,275 | 21,538 |
| | India | 390,796 | 208,674 | 101,221 | 60,970 | 19,932 |
| | Iran | 939,551 | 269,211 | 387,750 | 213,728 | 68,862 |
| | Iraq | 1,067,101 | 465,979 | 406,507 | 140,485 | 54,130 |
| | Kyrgyzstan | 267,871 | 21,547 | 49,269 | 170,328 | 26,727 |
| | Malaysia | 259,768 | 88,759 | 136,172 | 9,465 | 23,271 |
| | Nepal | 756,865 | 173,549 | 298,966 | 186,779 | 97,571 |
| | Philippines | 4,434,302 | 805,703 | 1,578,499 | 1,634,191 | 292,759 |
| | Thailand | 449,868 | 204,711 | 156,584 | 68,491 | 20,081 |
| | Vietnam | 10,419,190 | 2,944,924 | 5,369,445 | 1,471,772 | 632,544 |
| Europe (n=8) | Belarus | 676,888 | 2,669 | 193,952 | 389,202 | 85,950 |
| | France | 15,973,798 | 2,386,621 | 3,988,739 | 6,167,221 | 3,431,216 |
| | Greece | 844,332 | 94,397 | 363,723 | 269,670 | 116,542 |
| | Portugal | 440,013 | 166,749 | 139,831 | 72,654 | 60,779 |
| | Romania | 1,622,526 | 201,115 | 466,675 | 717,013 | 237,723 |
| | Slovenia | 140,192 | 9,077 | 36,064 | 84,082 | 10,969 |
| | Spain | 3,316,444 | 369,054 | 1,451,175 | 1,100,323 | 395,891 |
| | Switzerland | 279,943 | 11,757 | na | 230,720 | 19,796 |
| **Latin America (n=17)** | Argentina | 2,586,143 | 409,493 | 1,301,578 | 730,239 | 144,833 |
| | Bolivia | 502,284 | 168,163 | 192,031 | 116,704 | 19,380 |
| | Brazil | 7,235,880 | 2,537,580 | 2,158,681 | 1,870,087 | 669,532 |
| | Chile | 1,085,157 | 181,395 | 483,372 | 371,705 | 48,685 |
| | Colombia | 2,656,151 | 687,442 | 1,012,919 | 712,921 | 223,381 |
| | Costa Rica | 315,966 | 52,954 | 153,697 | 57,115 | 52,200 |
| | Cuba | 832,331 | 73,841 | 421,343 | 268,206 | 68,941 |
| | Dominican Republic | 640,133 | 198,585 | 239,538 | 148,380 | 53,630 |
| | Ecuador | 981,329 | 190,138 | 417,162 | 268,511 | 74,062 |
| | El Salvador | 370,047 | 159,942 | 134,542 | 61,548 | 14,015 |
| | Haiti | 531,559 | 319,866 | 148,872 | 59,137 | 3,684 |
| | Jamaica | 128,461 | 6,531 | 42,647 | 71,757 | 2,356 |
| | Nicaragua | 320,999 | 149,034 | 106,765 | 46,829 | 17,061 |
| | Paraguay | 325,045 | 110,471 | 146,315 | 52,009 | 12,383 |
| | Peru | 1,902,592 | 473,617 | 422,957 | 821,756 | 184,262 |
| | Uruguay | 245,345 | 28,219 | 145,157 | 58,472 | 13,497 |
| | Trinidad and Tobago | 82,872 | 12,863 | 28,729 | 37,372 | 1,885 |
| **North America (n=3)** | Canada | 623,869 | 13,536 | 184,551 | 297,787 | 127,995 |
| | United States | 10,718,751 | 198,623 | 1,968,655 | 6,308,388 | 2,243,085 |
| | Mexico | 7,974,041 | 1,620,961 | 4,036,280 | 1,526,107 | 761,884 |

*Note: Unweighted sample sizes, population 15 years and over. Summing numbers with different levels of educational attainment does not correspond to the recorded total because educational attainment is not known for some respondents.*

**Table A4.1** Crude migration intensities by educational attainment in Africa, Asia and Europe

| Region | country | Administrative units (regions) | Total population | Less than primary | Primary completed | Secondary completed | University completed |
|---|---|---|---|---|---|---|---|
| Africa | Botswana | major | 22.32 | 7.98 | 21.04 | 34.69 | 41.33 |
| | Cameroon | major | 8.83 | 4.40 | 10.15 | 16.78 | 16.21 |
| | | minor | 12.95 | 6.66 | 14.98 | 23.39 | 21.86 |
| | Egypt | major | 1.65 | 1.26 | 1.25 | 1.99 | 2.65 |
| | Ethiopia | major | 9.06 | 6.22 | 19.31 | 28.03 | 34.79 |
| | Ghana | major | 3.97 | 2.90 | 4.53 | 6.44 | 5.82 |
| | | minor | 6.62 | 4.63 | 7.78 | 10.79 | 10.34 |
| | Guinea | major | 13.65 | 11.99 | 23.87 | 32.59 | 36.16 |
| | | minor | 5.96 | 6.03 | 6.12 | 5.31 | 3.31 |
| | Kenya | major | 16.61 | 7.62 | 16.40 | 29.12 | 40.65 |
| | Malawi | major | 10.63 | 7.17 | 14.61 | 26.71 | 41.80 |
| | Mali | major | 6.74 | 5.64 | 9.83 | 19.01 | 17.71 |
| | | minor | 7.98 | 6.82 | 11.35 | 20.88 | 18.99 |
| | Morocco | major | 5.29 | 1.65 | 5.44 | 7.98 | 17.40 |
| | | minor | 8.83 | 3.99 | 9.21 | 13.10 | 23.51 |
| | Mozambique | major | 2.88 | 2.06 | 6.10 | 11.70 | 15.81 |
| | | minor | 7.63 | 6.28 | 13.28 | 20.28 | 19.45 |
| | Rwanda | minor | 13.77 | 11.23 | 18.36 | 43.91 | 68.72 |
| | Senegal | major | 4.29 | 3.62 | 5.56 | 9.23 | 8.18 |
| | | minor | 8.63 | 7.25 | 11.70 | 16.60 | 15.48 |
| | South Africa | major | 4.67 | 2.46 | 4.20 | 7.58 | 10.80 |
| | Uganda | minor | 8.79 | 6.23 | 10.86 | 17.67 | 24.44 |
| Asia | Armenia | major | 2.79 | 1.01 | 1.79 | 2.71 | 3.69 |
| | Cambodia | major | 10.34 | 8.14 | 11.64 | 18.83 | 30.49 |
| | China | major | 12.30 | 4.27 | 11.03 | 24.45 | 35.19 |
| | Fiji | major | 15.58 | 8.73 | 13.56 | 20.97 | 23.86 |
| | India | major | 1.21 | 0.83 | 1.33 | 1.78 | 2.84 |
| | India | minor | 3.92 | 2.65 | 4.05 | 6.06 | 9.92 |
| | Indonesia | major | 2.18 | 0.64 | 1.73 | 4.41 | 5.66 |
| | | minor | 4.39 | 1.38 | 3.57 | 8.59 | 11.60 |
| | Iran | major | 3.99 | 1.62 | 4.05 | 5.78 | 7.42 |
| | | minor | 6.84 | 2.94 | 6.85 | 9.84 | 12.75 |
| | Iraq | major | 2.88 | 2.83 | 2.92 | 2.84 | 3.21 |
| | Kyrgyzstan | major | 8.12 | 2.37 | 5.35 | 9.26 | 10.64 |
| | | minor | 11.39 | 3.60 | 7.48 | 13.07 | 14.19 |
| | Malaysia | major | 5.29 | 1.65 | 5.44 | 7.98 | 17.40 |
| | | minor | 8.83 | 3.99 | 9.21 | 13.10 | 23.51 |
| | Nepal | major | 6.09 | 3.90 | 4.09 | 5.20 | 11.20 |
| | Philippines | major | 3.62 | 1.76 | 3.05 | 4.80 | 5.03 |
| | | minor | 4.97 | 2.68 | 4.24 | 6.39 | 7.08 |
| | Thailand | major | 4.39 | 1.90 | 5.33 | 8.51 | 8.32 |
| | Vietnam | major | 5.00 | 2.18 | 4.82 | 9.71 | 8.79 |
| | | minor | 7.41 | 3.20 | 6.84 | 13.82 | 16.99 |
| Europe | Belarus | major | 6.85 | 4.38 | 4.39 | 8.01 | 7.47 |
| | France | minor | 10.44 | 4.48 | 5.89 | 10.56 | 19.64 |
| | | major | 6.62 | 2.69 | 3.90 | 6.60 | 12.55 |
| | Greece | minor | 10.39 | 3.70 | 6.86 | 14.40 | 17.53 |
| | | major | 6.17 | 2.34 | 3.88 | 8.82 | 10.27 |
| | Portugal | major | 2.74 | 1.29 | 2.36 | 3.67 | 6.49 |
| | | minor | 6.79 | 2.88 | 6.57 | 9.79 | 14.46 |
| | Romania | major | 2.12 | 0.81 | 1.31 | 1.96 | 5.29 |
| | Slovenia | major | 2.26 | 1.18 | 1.28 | 2.29 | 6.19 |
| | | minor | 7.27 | 2.99 | 4.81 | 7.97 | 13.47 |
| | Spain | major | 3.75 | 2.10 | 2.75 | 4.59 | 6.60 |
| | | minor | 10.72 | 4.99 | 8.10 | 13.88 | 16.92 |
| | Switzerland | major | 6.34 | 3.92 | - | 6.22 | 11.89 |
| | | minor | 20.51 | 14.38 | - | 20.73 | 28.07 |

*Note: results for population 15 years and over*

**Table A4.2** Crude migration intensities by educational attainment in Latin America and the Caribbean and North America

| Region | country | Administrative units (regions) | Total population | Less than primary | Primary completed | Secondary completed | University completed |
|---|---|---|---|---|---|---|---|
| Latin America and the Caribbean | Argentina | minor | 7.39 | 6.28 | 6.47 | 8.83 | 11.54 |
| | | major | 3.82 | 2.36 | 3.05 | 5.21 | 7.95 |
| | Bolivia | minor | 12.48 | 9.48 | 13.60 | 14.95 | 13.10 |
| | | major | 7.36 | 5.02 | 7.79 | 9.60 | 10.05 |
| | Brazil | minor | 7.59 | 5.91 | 8.16 | 8.56 | 9.44 |
| | | major | 2.67 | 1.98 | 2.88 | 2.98 | 3.68 |
| | Chile | minor | 17.37 | 10.06 | 15.56 | 21.79 | 28.95 |
| | | major | 7.58 | 4.39 | 6.73 | 9.64 | 12.19 |
| | Colombia | minor | 7.59 | 5.79 | 7.54 | 8.96 | 8.87 |
| | | major | 4.34 | 2.81 | 4.30 | 5.49 | 5.86 |
| | Costa Rica | minor | 10.63 | 8.59 | 9.58 | 11.57 | 14.74 |
| | | major | 5.54 | 4.43 | 4.98 | 6.14 | 7.69 |
| | Cuba | minor | 5.71 | 3.43 | 5.08 | 6.63 | 8.48 |
| | | major | 2.61 | 1.69 | 2.58 | 2.78 | 3.11 |
| | Dominican Republic | minor | 6.24 | 5.16 | 6.52 | 7.34 | 5.97 |
| | | major | 5.13 | 4.21 | 5.30 | 6.13 | 4.95 |
| | Ecuador | major | 5.22 | 3.91 | 5.23 | 6.47 | 5.26 |
| | El Salvador | minor | 5.87 | 4.55 | 6.47 | 7.48 | 8.10 |
| | | major | 3.26 | 2.58 | 3.51 | 4.02 | 5.15 |
| | Haiti | major | 2.41 | 2.07 | 2.95 | 2.95 | 2.12 |
| | Jamaica | major | 8.87 | 3.73 | 4.50 | 11.82 | 14.50 |
| | Nicaragua | minor | 4.61 | 4.80 | 4.42 | 4.23 | 5.12 |
| | | major | 2.87 | 2.82 | 2.75 | 2.99 | 3.69 |
| | Paraguay | minor | 14.94 | 11.42 | 15.61 | 17.11 | 15.79 |
| | | major | 10.87 | 7.62 | 11.18 | 12.91 | 13.34 |
| | Peru | major | 6.87 | 3.96 | 6.79 | 8.29 | 8.21 |
| | | minor | 9.45 | 5.92 | 9.44 | 11.11 | 11.13 |
| | Trinidad and Tobago | major | 8.91 | 5.15 | 7.15 | 10.77 | 20.27 |
| | Uruguay | major | 5.14 | 2.75 | 4.44 | 7.70 | 6.53 |
| | | minor | 7.28 | 4.92 | 6.92 | 9.31 | 7.36 |
| North America | Canada | minor | 12.42 | 5.75 | 8.66 | 12.90 | 17.42 |
| | | major | 3.34 | 0.97 | 2.22 | 3.29 | 5.34 |
| | Mexico | major | 3.60 | 1.86 | 3.38 | 4.76 | 6.12 |
| | | minor | 7.03 | 3.57 | 6.69 | 9.22 | 11.77 |
| | United States | major | 9.00 | 4.84 | 5.81 | 8.43 | 13.79 |

*Note: results for population 15 years and over*

**Table A5.1** Size of migration flows by educational attainment in Africa, Asia and Europe

| Region | country | Administrative units (regions) | Total population | Less than primary | Primary completed | Secondary completed | University completed |
|---|---|---|---|---|---|---|---|
| Africa | Botswana | major | 266,443 | 20,202 | 129,840 | 85,001 | 29,439 |
| | Cameroon | major | 860,903 | 152,130 | 509,516 | 147,166 | 36,314 |
| | | minor | 1,262,593 | 230,270 | 751,975 | 205,137 | 48,971 |
| | Egypt | major | 818,121 | 257,041 | 93,314 | 322,801 | 144,460 |
| | Ethiopia | major | 991,266 | 592,845 | 216,235 | 73,748 | 9,373 |
| | Ghana | major | 439,947 | 160,037 | 183,499 | 88,268 | 8,259 |
| | | minor | 733,615 | 255,507 | 315,149 | 147,890 | 14,672 |
| | Guinea | major | 145,008 | 88,073 | 17,038 | 5,446 | 1,751 |
| | | minor | 498,579 | 364,137 | 89,780 | 27,558 | 10,881 |
| | Kenya | major | 3,579,002 | 524,463 | 1,614,875 | 1,307,939 | 132,604 |
| | Malawi | major | 761,952 | 327,721 | 240,400 | 159,718 | 11,541 |
| | Mali | major | 456,923 | 312,822 | 95,403 | 31,416 | 17,177 |
| | | minor | 540,986 | 378,271 | 110,155 | 34,506 | 18,418 |
| | Morocco | major | 1,056,000 | 651,740 | 240,580 | 114,360 | 49,320 |
| | | minor | 1,441,560 | 897,100 | 325,360 | 153,840 | 65,260 |
| | Mozambique | major | 308,498 | 179,077 | 95,380 | 23,388 | 4,392 |
| | | minor | 817,304 | 545,924 | 207,647 | 40,540 | 5,403 |
| | Rwanda | minor | 596,614 | 357,551 | 192,319 | 40,296 | 6,570 |
| | Senegal | major | 242,178 | 151,643 | 65,220 | 21,520 | 4,089 |
| | | minor | 487,179 | 303,704 | 137,243 | 38,703 | 7,738 |
| | South Africa | major | 1,153,607 | 162,622 | 496,541 | 429,525 | 65,111 |
| | Uganda | minor | 1,102,637 | 426,546 | 537,214 | 116,763 | 22,008 |
| Asia | Armenia | major | 61,962 | 682 | 3,781 | 39,213 | 18,270 |
| | Cambodia | major | 906,647 | 398,387 | 381,586 | 88,595 | 38,244 |
| | China | major | 111,011,424 | 6,116,489 | 67,380,847 | 33,553,835 | 3,929,667 |
| | Fiji | major | 90,651 | 3,621 | 48,706 | 35,362 | 2,980 |
| | India | major | 11,821,577 | 4,329,977 | 3,365,585 | 2,713,167 | 1,415,150 |
| | | minor | 38,298,002 | 13,824,626 | 10,248,586 | 9,236,963 | 4,943,059 |
| | Indonesia | major | 3,840,818 | 272,422 | 1,543,909 | 1,736,056 | 283,497 |
| | | minor | 7,734,492 | 587,410 | 3,185,985 | 3,381,569 | 581,020 |
| | Iran | major | 1,874,405 | 218,061 | 785,193 | 617,674 | 255,480 |
| | | minor | 3,213,265 | 395,740 | 1,328,042 | 1,051,542 | 438,998 |
| | Iraq | major | 307,325 | 131,872 | 118,700 | 39,898 | 17,376 |
| | Kyrgyzstan | major | 217,511 | 5,107 | 26,359 | 157,724 | 28,438 |
| | | minor | 305,105 | 7,757 | 36,853 | 222,619 | 37,926 |
| | Malaysia | major | 687,086 | 73,226 | 370,388 | 37,765 | 202,458 |
| | | minor | 1,146,876 | 177,074 | 627,072 | 61,996 | 273,551 |
| | Nepal | major | 921,862 | 135,368 | 244,554 | 194,250 | 218,559 |
| | Philippines | major | 1,605,217 | 141,804 | 481,442 | 784,412 | 147,258 |
| | | minor | 2,203,848 | 215,928 | 669,284 | 1,044,248 | 207,273 |
| | Thailand | major | 1,974,918 | 388,950 | 834,594 | 582,863 | 167,075 |
| | Vietnam | major | 3,473,063 | 427,996 | 1,725,381 | 952,727 | 370,670 |
| | | minor | 5,147,080 | 628,250 | 2,448,467 | 1,355,993 | 716,461 |
| Europe | Belarus | major | 463,668 | 1,169 | 85,145 | 311,751 | 64,205 |
| | France | minor | 5,053,529 | 324,002 | 711,929 | 1,973,511 | 2,042,094 |
| | | major | 3,204,441 | 194,546 | 471,396 | 1,233,444 | 1,304,902 |
| | Greece | minor | 877,261 | 34,927 | 249,514 | 388,325 | 204,298 |
| | | major | 520,953 | 22,089 | 141,125 | 237,849 | 119,689 |
| | Portugal | major | 241,127 | 43,021 | 66,000 | 53,328 | 78,891 |
| | | minor | 597,538 | 96,047 | 183,738 | 142,257 | 175,773 |
| | Romania | major | 76,534 | 24,829 | 35,624 | 36,582 | 4,494 |
| | Slovenia | major | 31,683 | 1,071 | 4,616 | 19,255 | 6,790 |
| | | minor | 101,920 | 2,714 | 17,347 | 67,013 | 14,775 |
| | Spain | major | 1,243,666 | 77,501 | 399,073 | 505,048 | 261,288 |
| | | minor | 3,555,228 | 184,158 | 1,175,452 | 1,527,248 | 669,848 |
| | Switzerland | major | 354,968 | 9,217 | - | 287,016 | 47,075 |
| | | minor | 1,148,326 | 33,813 | - | 956,565 | 111,135 |

*Note: The sum of migrants with different educational attainment does not add up to the total number of migrants because educational attainment is not known for some respondents. The number of migrants between minor regions includes those who move between major regions so these two figures should not be summed.*

**Table A5.2** Size of migration flows by educational attainment in Latin America and the Caribbean and North America

| Region | country | Administrative units (regions) | Total population | Less than primary | Primary completed | Secondary completed | University completed |
|---|---|---|---|---|---|---|---|
| Latin America and the Caribbean | Argentina | minor | 1,911,160 | 257,162 | 842,121 | 644,801 | 167,137 |
| | | major | 987,907 | 96,640 | 396,981 | 380,455 | 115,142 |
| | Bolivia | minor | 626,850 | 159,419 | 261,162 | 174,472 | 25,388 |
| | | major | 369,681 | 84,418 | 149,592 | 112,036 | 19,477 |
| | Brazil | minor | 10,984,066 | 2,999,420 | 3,522,967 | 3,201,589 | 1,264,076 |
| | | major | 3,863,960 | 1,004,882 | 1,243,400 | 1,114,572 | 492,775 |
| | Chile | minor | 1,884,918 | 182,483 | 752,127 | 809,945 | 140,943 |
| | | major | 822,549 | 79,632 | 325,309 | 358,324 | 59,347 |
| | Colombia | minor | 2,016,018 | 398,029 | 763,741 | 638,777 | 198,139 |
| | | major | 1,152,769 | 193,171 | 435,555 | 391,394 | 130,901 |
| | Costa Rica | minor | 335,872 | 45,487 | 147,242 | 66,082 | 76,943 |
| | | major | 175,045 | 23,459 | 76,541 | 35,069 | 40,142 |
| | Cuba | minor | 475,261 | 25,327 | 214,042 | 177,821 | 58,462 |
| | | major | 217,238 | 12,479 | 108,706 | 74,561 | 21,441 |
| | Dominican Republic | minor | 399,443 | 102,470 | 156,179 | 108,911 | 32,017 |
| | | major | 328,388 | 83,604 | 126,955 | 90,957 | 26,547 |
| | Ecuador | major | 512,254 | 74,344 | 218,176 | 173,727 | 38,957 |
| | El Salvador | minor | 217,218 | 72,774 | 87,049 | 46,038 | 11,352 |
| | | major | 120,635 | 41,265 | 47,224 | 24,742 | 7,218 |
| | Haiti | major | 128,106 | 66,212 | 43,917 | 17,445 | 781 |
| | Jamaica | major | 113,945 | 2,436 | 19,191 | 84,817 | 3,416 |
| | Nicaragua | minor | 147,981 | 71,536 | 47,190 | 19,809 | 8,735 |
| | | major | 92,127 | 42,028 | 29,360 | 14,002 | 6,296 |
| | Paraguay | minor | 485,617 | 126,158 | 228,398 | 88,988 | 19,552 |
| | | major | 353,324 | 84,179 | 163,580 | 67,144 | 16,518 |
| | Peru | major | 1,307,081 | 187,552 | 287,188 | 681,236 | 151,279 |
| | | minor | 1,797,949 | 280,381 | 399,271 | 912,971 | 205,084 |
| | Trinidad and Tobago | major | 73,839 | 6,624 | 20,541 | 40,250 | 3,821 |
| | Uruguay | major | 126,107 | 7,760 | 64,450 | 45,023 | 8,814 |
| | | minor | 178,611 | 13,884 | 100,449 | 54,437 | 9,934 |
| North America | Canada | minor | 2,869,797 | 28,826 | 591,930 | 1,422,762 | 825,802 |
| | | major | 771,749 | 4,863 | 151,742 | 362,860 | 253,145 |
| | Mexico | major | 2,708,165 | 284,433 | 1,287,040 | 685,309 | 439,880 |
| | | minor | 5,288,444 | 545,928 | 2,547,426 | 1,327,425 | 845,979 |
| | United States | major | 19,293,751 | 192,267 | 2,287,577 | 10,635,941 | 6,186,429 |

*Note: The sum of migrants with different educational attainment does not add up to the total number of migrants because educational attainment is not known for some respondents. The number of migrants between minor regions includes those who move between major regions so these two figures should not be summed.*

**Table A6.1** Age and migration intensity at peak in Africa, Asia and Europe

| | Country | Administrative units (regions) | age at peak | Intensity at peak |
|---|---|---|---|---|
| Africa | Botswana | major | 22.0 | 0.0307 |
| | Cameroon | major | 23.0 | 0.0324 |
| | | minor | 22.5 | 0.0308 |
| | Egypt | major | 27.5 | 0.0325 |
| | Ethiopia | major | 20.5 | 0.0356 |
| | Ghana | major | 23.5 | 0.0267 |
| | | minor | 24.0 | 0.0239 |
| | Guinea | major | 23.0 | 0.0307 |
| | Kenya | minor | 23.5 | 0.0360 |
| | Malawi | major | 24.5 | 0.0284 |
| | Mali | major | 22.0 | 0.0312 |
| | | minor | 22.0 | 0.0303 |
| | Morocco | major | 22.0 | 0.0410 |
| | | minor | 22.5 | 0.0364 |
| | Mozambique | major | 23.0 | 0.0318 |
| | | minor | 22.0 | 0.0260 |
| | Rwanda | minor | 24.5 | 0.0288 |
| | Senegal | major | 23.0 | 0.0277 |
| | | minor | 22.5 | 0.0234 |
| | South Africa | major | 24.5 | 0.0340 |
| | Uganda | minor | 22.5 | 0.0324 |
| Asia | Armenia | major | 23.5 | 0.0385 |
| | Cambodia | major | 23.0 | 0.0351 |
| | | minor | 23.0 | 0.0329 |
| | China | major | 20.5 | 0.0457 |
| | Fiji | major | 21.5 | 0.0265 |
| | India | major | 22.0 | 0.0463 |
| | | minor | 22.0 | 0.0492 |
| | Indonesia | major | 22.5 | 0.0432 |
| | | minor | 22.5 | 0.0428 |
| | Iran | major | 24.5 | 0.0455 |
| | | minor | 24.5 | 0.0391 |
| | Iraq | major | 23.5 | 0.0224 |
| | Kyrgyzstan | major | 20.0 | 0.0371 |
| | | minor | 20.0 | 0.0355 |
| | Malaysia | major | 22.0 | 0.0410 |
| | | minor | 22.5 | 0.0364 |
| | Nepal | major | 21.5 | 0.0398 |
| | Philippines | major | 24.5 | 0.0297 |
| | | minor | 24.5 | 0.0290 |
| | Thailand | major | 23.5 | 0.0411 |
| | Vietnam | major | 20.5 | 0.0620 |
| | | minor | 21.0 | 0.0538 |
| Europe | Belarus | major | 19.5 | 0.0562 |
| | France | major | 26.0 | 0.0367 |
| | | minor | 26.5 | 0.0362 |
| | Greece | major | 21.0 | 0.0301 |
| | | minor | 28.0 | 0.0257 |
| | Portugal | major | 29.0 | 0.0420 |
| | | minor | 29.5 | 0.0386 |
| | Romania | major | 24.0 | 0.0450 |
| | Spain | major | 30.0 | 0.0350 |
| | | minor | 31.5 | 0.0363 |
| | Switzerland | major | 25.5 | 0.0390 |
| | | minor | 27.0 | 0.0356 |

*Note: single year age groups between 5 and 65, data was smoothed using kernel regression and normalised to unity*

**Table A6.2** Age and migration intensity at peak in Latin American and the Caribbean and North America

| | | | | |
|---|---|---|---|---|
| Latin America and the Caribbean | Argentina | between major | 27.5 | 0.0274 |
| | | between minor | 27.5 | 0.0259 |
| | Bolivia | between major | 20.5 | 0.0301 |
| | | between minor | 20.5 | 0.0284 |
| | Brazil | between major | 23.5 | 0.0284 |
| | | between minor | 23.5 | 0.0257 |
| | Chile | between major | 21.5 | 0.0289 |
| | | between minor | 28.0 | 0.0259 |
| | Columbia | between major | 21.0 | 0.0274 |
| | | between minor | 21.0 | 0.0253 |
| | Costa Rica | between major | 28.5 | 0.0245 |
| | | between minor | 29.5 | 0.0251 |
| | Cuba | between major | 24.0 | 0.0257 |
| | | between minor | 25.5 | 0.0241 |
| | Dominican Republic | between major | 22.0 | 0.0267 |
| | | between minor | 22.0 | 0.0254 |
| | Ecuador | between major | 22.0 | 0.0294 |
| | El Salvador | between major | 25.5 | 0.0265 |
| | | between minor | 26.0 | 0.0261 |
| | Haiti | between major | 21.5 | 0.0310 |
| | | between minor | 21.0 | 0.0217 |
| | Jamaica | between major | 20.0 | 0.0468 |
| | Nicaragua | between major | 21.0 | 0.0232 |
| | | between minor | 22.5 | 0.0225 |
| | Paraguay | between major | 23.5 | 0.0262 |
| | | between minor | 24.0 | 0.0263 |
| | Peru | between major | 20.0 | 0.0315 |
| | | between minor | 20.0 | 0.0306 |
| | Trinidad and Tobago | between major | 27.0 | 0.0282 |
| | Uruguay | between major | 21.0 | 0.0340 |
| | | between minor | 21.0 | 0.0309 |
| North America | Canada | between major | 26.5 | 0.0349 |
| | | between minor | 26.5 | 0.0330 |
| | Mexico | between major | 27.0 | 0.0264 |
| | | between minor | 28.0 | 0.0265 |
| | USA | between major | 25.5 | 0.0308 |

*Note: single year age groups between 5 and 65, data was smoothed using kernel regression and normalised to unity*

**Table A7** Mean years of schooling by migrant status, individuals aged 15 years and over

| | | Urban in-migrant | Rural in-migrant | Urban Non-migrant | Rural Non-migrant | Total |
|---|---|---|---|---|---|---|
| Africa | Cameroon | 7.93 | 4.97 | 6.61 | 3.27 | 5.02 |
| | Ghana | 7.37 | 4.53 | 5.84 | 3.21 | 4.44 |
| | Guinea | 3.10 | 1.68 | 3.33 | 0.52 | 1.42 |
| | Kenya | 6.61 | 2.36 | 6.81 | 4.51 | 4.98 |
| | Malawi | 6.53 | 3.09 | 6.42 | 3.51 | 3.90 |
| | Mali | 3.75 | 2.23 | 3.65 | 0.91 | 1.54 |
| | Rwanda | 5.38 | 2.74 | 3.77 | 2.52 | 2.83 |
| | Senegal | 4.43 | 1.87 | 3.95 | 1.00 | 2.25 |
| | South Africa | 8.92 | 6.64 | 7.59 | 4.98 | 6.53 |
| | Uganda | 7.10 | 4.53 | 6.21 | 3.43 | 3.86 |
| Asia | Cambodia | 6.38 | 3.99 | 5.58 | 3.39 | 3.84 |
| | Fiji | 9.31 | 8.65 | 8.69 | 7.66 | 8.29 |
| | Indonesia | 9.89 | 7.56 | 7.89 | 5.60 | 6.63 |
| | Thailand | 8.78 | 7.08 | 7.25 | 5.20 | 5.88 |
| | Vietnam | 10.27 | 8.66 | 8.59 | 6.83 | 7.44 |
| Latin America | Argentina | 9.38 | 7.08 | 7.93 | 5.47 | 7.43 |
| | Bolivia | 8.00 | 6.00 | 7.41 | 3.87 | 6.01 |
| | Chile | 9.95 | 8.27 | 8.76 | 6.09 | 8.52 |
| | Columbia | 8.10 | 5.10 | 7.35 | 3.99 | 6.60 |
| | Costa Rica | 9.03 | 6.82 | 8.12 | 6.28 | 7.65 |
| | Dominican Republic | 7.95 | 5.93 | 7.45 | 5.36 | 6.94 |
| | Ecuador | 8.83 | 7.24 | 9.97 | 7.69 | 8.33 |
| | El Salvador | 7.12 | 4.17 | 6.52 | 3.64 | 5.46 |
| | Haiti | 4.98 | 2.46 | 5.53 | 2.14 | 3.54 |
| | Jamaica | 10.68 | 9.10 | 9.80 | 8.79 | 9.34 |
| | Mexico | 9.08 | 6.48 | 8.01 | 5.21 | 7.40 |
| | Nicaragua | 6.77 | 3.41 | 6.47 | 3.02 | 4.90 |
| | Paraguay | 7.54 | 5.19 | 6.95 | 4.45 | 5.94 |
| | Peru | 8.86 | 7.03 | 8.18 | 4.73 | 7.42 |

*Note: results for population 15 years and over, migration between major regions*

**Table A8** Mean years of schooling by migrant status, 20 to 24 years old

|  |  | Urban in-migrant | Rural in-migrant | Urban Non-migrant | Rural Non-migrant |
|---|---|---|---|---|---|
| Africa | Cameroon | 9.97 | 6.65 | 9.35 | 4.73 |
| | Ghana | 8.69 | 5.81 | 7.77 | 4.66 |
| | Guinea | 3.17 | 1.36 | 4.93 | 0.49 |
| | Kenya | 9.78 | 7.95 | 8.97 | 6.83 |
| | Malawi | 9.21 | 6.84 | 9.17 | 5.67 |
| | Mali | 5.69 | 2.89 | 5.68 | 1.99 |
| | Rwanda | 5.93 | 3.81 | 5.57 | 4.01 |
| | Senegal | 5.30 | 2.24 | 5.23 | 1.37 |
| | South Africa | 10.46 | 8.47 | 10.10 | 8.30 |
| | Uganda | 8.65 | 6.03 | 8.56 | 5.15 |
| Asia | Cambodia | 7.49 | 5.39 | 8.18 | 5.25 |
| | Fiji | 11.53 | 10.85 | 11.44 | 10.42 |
| | Indonesia | 11.50 | 9.44 | 10.65 | 8.20 |
| | Thailand | 10.80 | 8.92 | 10.87 | 8.43 |
| | Vietnam | 11.47 | 9.89 | 10.56 | 8.83 |
| Latin America | Argentina | 10.88 | 8.01 | 10.41 | 7.89 |
| | Bolivia | 9.86 | 8.01 | 10.85 | 6.27 |
| | Chile | 11.82 | 10.15 | 11.63 | 9.03 |
| | Columbia | 10.20 | 6.89 | 10.35 | 6.20 |
| | Costa Rica | 10.37 | 8.27 | 10.20 | 8.77 |
| | Dominican Republic | 10.75 | 8.71 | 10.46 | 8.54 |
| | Ecuador | 10.68 | 8.86 | 11.30 | 8.83 |
| | El Salvador | 9.16 | 6.37 | 9.64 | 6.29 |
| | Haiti | 7.18 | 4.90 | 8.35 | 4.41 |
| | Jamaica | 13.20 | 12.54 | 12.84 | 12.00 |
| | Mexico | 10.95 | 9.11 | 10.93 | 8.58 |
| | Nicaragua | 8.46 | 4.72 | 9.00 | 4.49 |
| | Paraguay | 9.26 | 7.10 | 9.95 | 6.76 |
| | Peru | 10.42 | 8.56 | 10.64 | 7.52 |

*Note: results for population 20 to 24 years, migration between major regions*

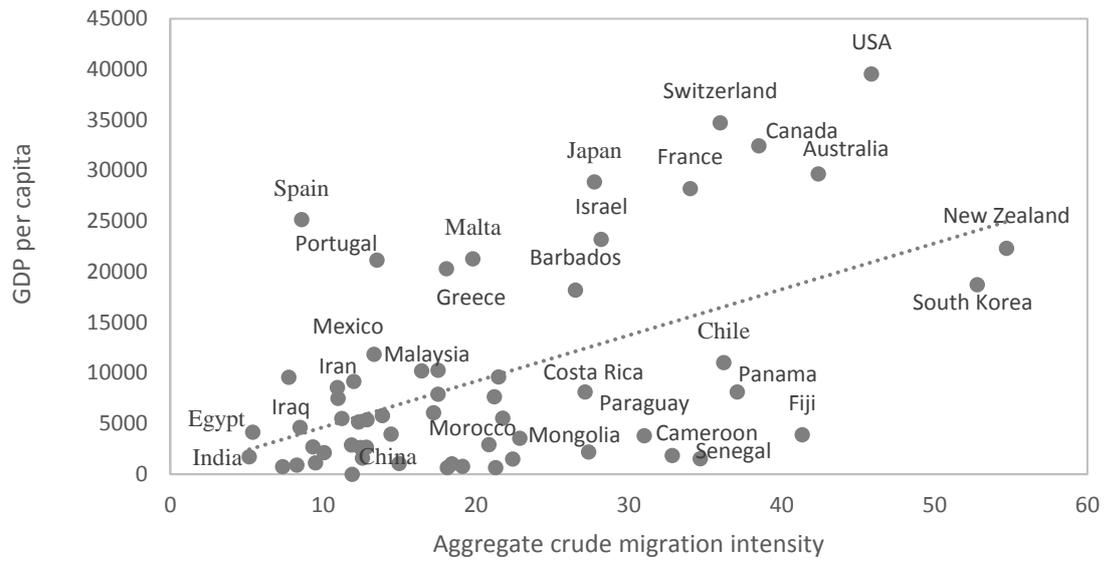

**Figure A1** Five-year aggregate crude migration intensity against HDI

*Note: GDP per capita measured as an average between 2000 and 2005*

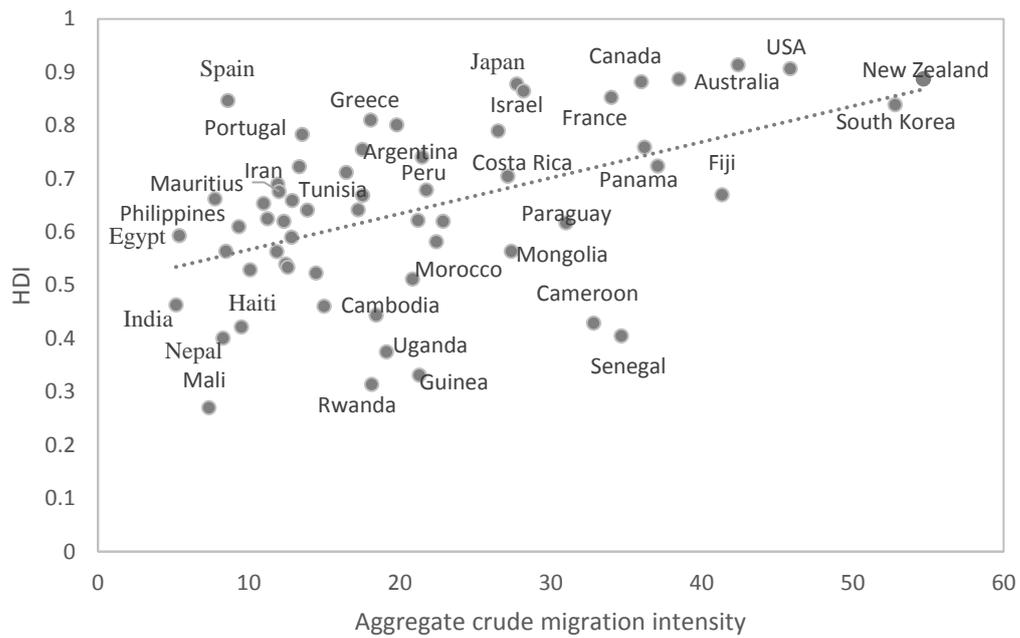

**Figure A2** Five-year aggregate crude migration intensity against GDP per capita

*Note: HDI measured as an average between 2000 and 2005*

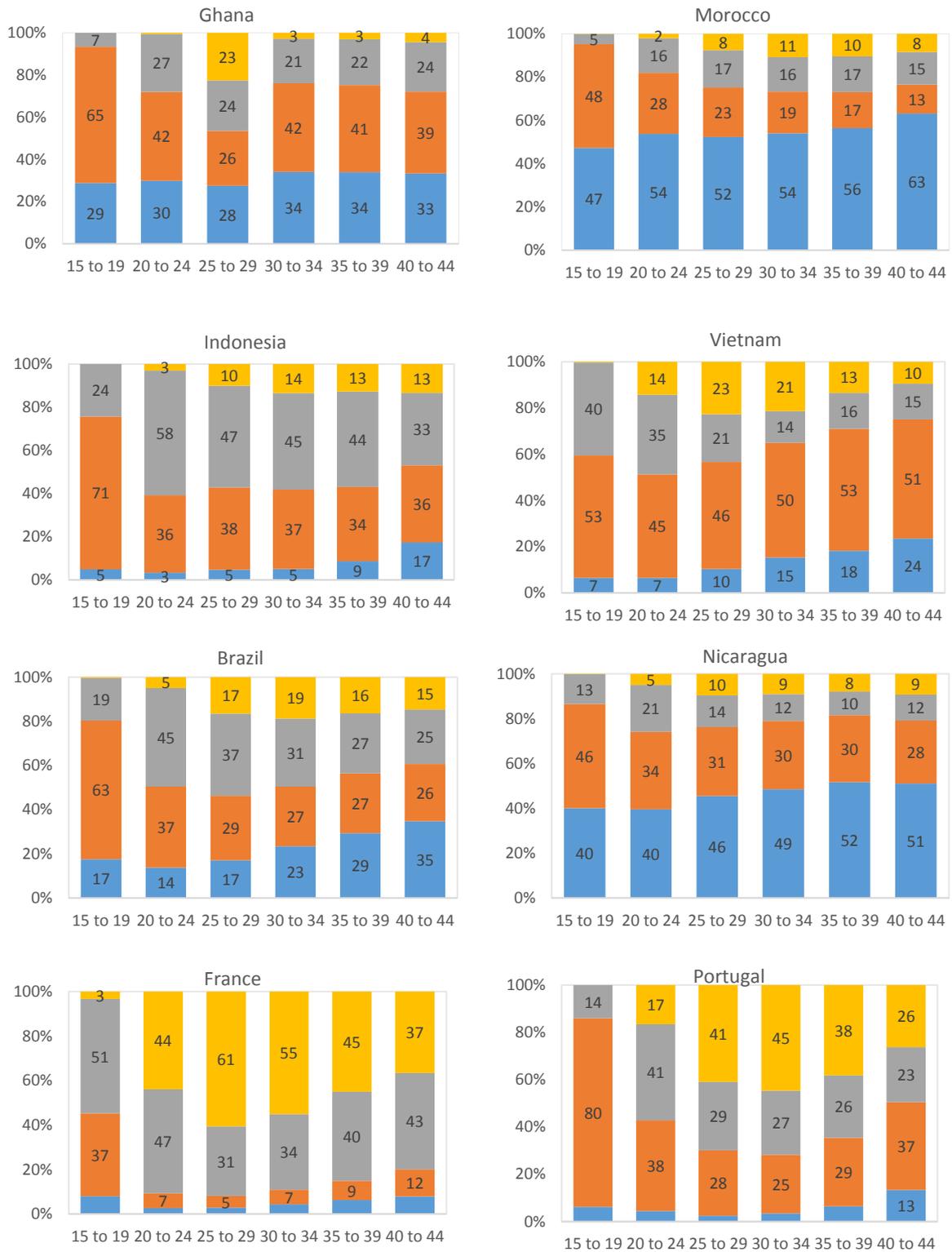

**Figure A3** Percentage distribution of migrants by age and educational attainment

*Note: migration between minor regions*

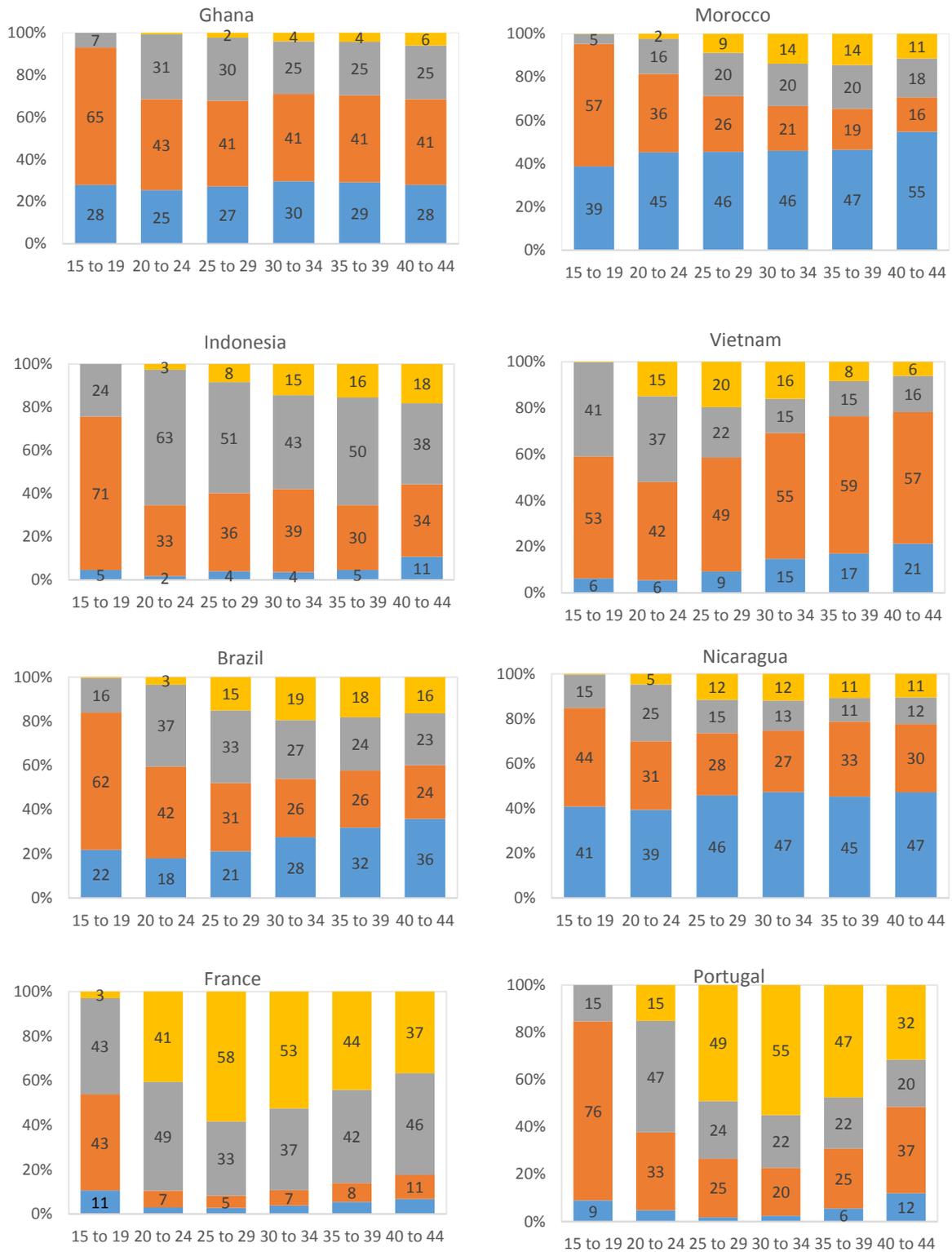

**Figure A4** Percentage distribution of migrants by age and educational attainment, males

*Note: migration between major regions*

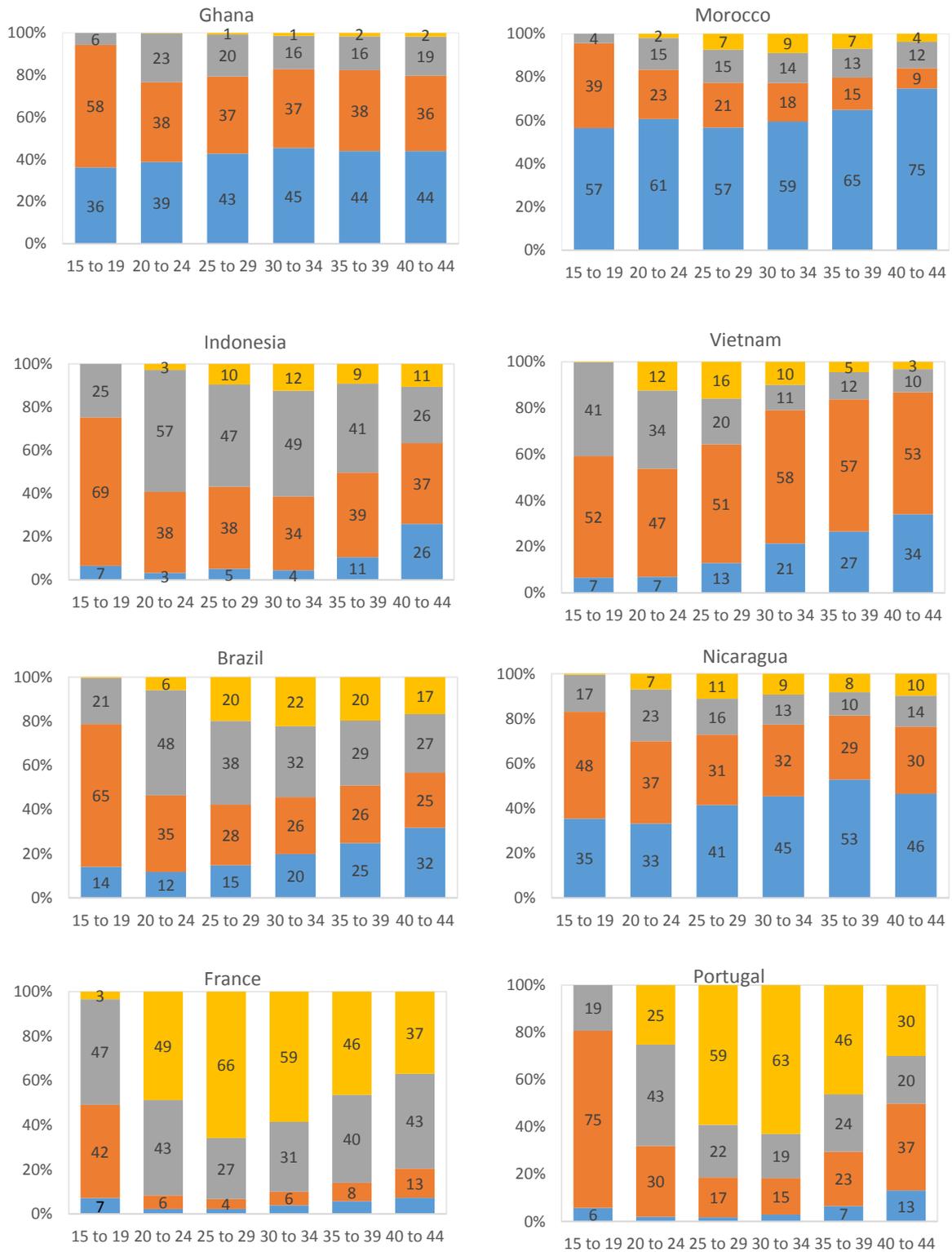

**Figure A4** Percentage distribution of migrants by age and educational attainment, females

*Note: migration between major regions*